\magnification=1200
\hoffset=-0.3 true cm
\hsize=15.5truecm
\vsize=23truecm
\catcode`\@=11
\def\gsim{\ifmmode{\mathrel{\mathpalette\@versim>}}
    \else{$\mathrel{\mathpalette\@versim>}$}\fi}
\def\lsim{\ifmmode{\mathrel{\mathpalette\@versim<}}
    \else{$\mathrel{\mathpalette\@versim<}$}\fi}
\def\@versim#1#2{\lower 2.9truept \vbox{\baselineskip 0pt \lineskip
    0.5truept \ialign{$\m@th#1\hfil##\hfil$\crcr#2\crcr\sim\crcr}}}
\catcode`\@=12
\def\ref{\par\noindent\hangindent=1truecm}
\def\yr-1{\hbox{${\rm yr}^{-1}$}}

\def\msun{\hbox{$M_\odot$}}
\def\mg2{\hbox{${\rm Mg}_2$}}

\def\vtsn{\hbox{$\vartheta_{\rm SN}$}}

\def\lx{\hbox{$L_{\rm X}$}}
\def\feh{\hbox{${\rm [Fe/H]}$}}

\def\log{\hbox{${\rm Log}\, $}}

\def\re{r_{\rm e}}
\def\ie{I_{\rm e}}

\def\lb{\hbox{$L_{\rm B}$}}

\def\ho{\hbox{$H_\circ$}}
\def\h50{\hbox{$\ho /50$}}

\def\lgrav{L^{\pm}_{\rm grav}}

\def\pn{\par\noindent}

\def\yr{hbox{\rm yr}}
\def\pn{\par\noindent}

\def\3/2{\hbox{${3\over 2}$}}

\def\lsun{\hbox{$L_\odot$}}
\def\lb{\hbox{$L_{\rm B}$}}
\def\lv{\hbox{$L_{\rm V}$}}
\def\lx{\hbox{$L_{\rm X}$}}

\def\lgrav-{\hbox{$L_{\rm grav}^{-}$}}

\def\msun{\hbox{$M_\odot$}}

\def\yr-1{\hbox{${\rm yr}^{-1}$}}

\def\klu{(Dordrecht: Kluwer), p. }
$\,$

\vskip 2 truecm
\centerline{\bf THE IRON DISCREPANCY}
\centerline{\bf IN ELLIPTICAL GALAXIES}
\centerline{\bf AFTER ASCA}
\vskip 2cm
\centerline{NOBUO ARIMOTO}
\centerline{Institute of Astronomy, University of Tokyo}
\centerline{Mitaka, Tokyo 181, Japan}
\vskip 0.6cm
\centerline{KYOKO MATSUSHITA}
\centerline{YUHRI ISHIMARU}
\centerline{Department of Astronomy, University of Tokyo}
\centerline{Bunkyo-ku, Tokyo 113, Japan}
\vskip 0.6cm
\centerline{TAKAYA OHASHI}
\centerline{Department of Physics, Tokyo Metropolitan University}
\centerline{Hachioji, Tokyo 192-03, Japan}
\vskip 0.6cm
\centerline{AND}
\vskip 0.6cm
\centerline{ALVIO RENZINI$^1$}
\centerline{European Southern Observatory}
\centerline{ Karl-Schwarzschild-Strasse 2, D-85748,
Garching bei M\"unchen, Germany}

\bigskip\noindent
\vskip 3.5 true cm
\pn
\vskip 0.5 truecm
\pn
$^1$ On leave from: Dipartimento di Astronomia, Universit\`a di Bologna.

\vfill\eject
\baselineskip=6truemm
\centerline{\bf ABSTRACT}
\bigskip
We present estimates for the iron content of the stellar
and diffused components of elliptical galaxies, as derived respectively
from integrated optical spectra and from {\it ASCA} X-ray observations. 
A macroscopic discrepancy emerges  between the expected iron abundances
in the hot interstellar medium (ISM) and what is indicated 
by the X-ray observations, especially when allowance is made for the 
current iron enrichment by Type Ia supernovae. 
This strong discrepancy, that in some extreme instances may be as large as a
factor of $\sim 20$, calls into question our current understanding of supernova
enrichment and chemical evolution of galaxies.
We discuss several astrophysical
implications of the inferred low iron abundances in the ISM, including
the chemical evolution of galaxies and cluster of galaxies, the
evolution of gas flows in ellipticals, and the heating of the
intracluster medium.
Some of the consequences appear hard to accept, and in the attempt
to avoid some of these difficulties we explore ways
of hiding or diluting iron in the ISM of ellipticals. 
None of these possibilities  appears 
astrophysically plausible, and we alternatively rise the question of the
reliability of iron-L line diagonostic tools that are currently used to
infer abundances from X-ray spectra. Various thin plasma emission
models are shown to give iron abundances that may differ
significantly, especially at low temperatures ($kT\lsim 1$ keV),
 when the iron-L complex is dominated by iron ions with still many  
bound electrons. From a collection of {\it ASCA} and other X-ray
observatory data, it is shown that current thin plasma codes tend to
give very low iron abundances when the temperature of the 
objects is below $\sim 1$ keV. Such objects include various types
of binary stars, supernova remnants, starburst galaxies, AGN's,
with the case of galaxy groups being especially well documented.
We conclude that -- besides rethinking the chemical evolution of
galaxies -- one should also  consider the possibility that
existing thin plasma models may incorporate inaccurate atomic
physics for the ions responsible for the iron-L complex.

\smallskip
\pn
{\it Subject headings:} Galaxies: intergalactic medium -- Galaxies: clustering
    -- Galaxies: Abundances -- Galaxies: Evolution -- X-rays: Galaxies --
     Stars: Supernovae: General
\bigskip
\vfill\eject

\centerline{\bf 1. INTRODUCTION}
\medskip
Stars in elliptical galaxies lose about one third of their mass while 
ascending the red giant branches and before dying as white dwarfs. 
The ISM of these galaxies is therefore continuously replenished of gas,
that is soon heated to X-ray temperatures by the collision of the red 
giant winds with the ISM, by supernova (SN) explosions, and -- in the 
case of inflows -- by the $PdV$ work due to gravitational compression. 
Hence the chemical composition of the ISM is expected to reflect the average 
composition of the stellar component as established 
at the time of its formation, then further enriched by SNs of Type Ia that 
currently explode in elliptical galaxies. This additional enrichment
is expected to be especially important for iron, as this is the main
nucleosynthetic contribution of SNIa's (Nomoto, Thielemann, \& Yokoi, 
1984). On these premises the iron abundance in the hot ISM is expected to be:

$$Z^{\rm Fe}_{\rm ISM}= <\! Z^{\rm Fe}_*\!> + 
    5\vartheta_{\rm SN}\left({M^{\rm Fe}
   \over 0.7\,\msun}\right)h^2_{50},
\eqno(1)$$
(Ciotti et al. 1991; Renzini et al. 1993, hereafter RCDP),
where $<\! Z^{\rm Fe}_*\!>$ is the average iron abundance of the stars
in units of the solar abundance, $h_{50} \equiv \h50$ the Hubble
constant in units of 50 km s$^{-1}$ per Mpc, and $\vartheta_{\rm SN}$ 
is the present rate of SNIa in ellipticals in units of the rate as 
estimated by Tammann (1982), or $2.2\times 10^{-13}$
SNs $\lsun^{-1}\yr-1$. Current estimates of the iron yield 
($M^{\rm Fe}$) per SNIa event cluster around 0.7 $\msun$ of iron 
(Nomoto et al. 1984; Shigeyama et al. 1992; Woosley \& Weaver 1994; 
Nomoto et al. 1996).

Note that increasing the Hubble constant has the effect of increasing 
the expected iron enrichment from SNIa's. This comes from the SN rate per 
unit galaxy luminosity increasing as the luminosity decreases 
with a shorter distance scale, while the actual number of
observed SN events clearly stays the same. However, part of this effect may be
compensated by the estimated iron yield per SNIa event being itself dependent
on the distance scale, as it is based on fitting the SN light-curve. Still,
this compensation cannot be complete, as at least 0.4 $\msun$ need to be
incinerated to nuclear statistical equilibrium (and eventually processed to
iron), in order to unbound the star and impart to the SN photosphere the
observed high velocity (Arnett, Branch, \& Wheeler 1985). When bearing in mind
that the $0.7\,\msun$ estimate of Nomoto et al. (1984) is based on the
assumption $h_{50}=1$, we can guess $M^{\rm Fe}=0.4\,\msun$ to be more 
appropriate for the choice $h_{50}=2$. This still implies a more
pronounced iron emrichment by SNIa's if a short distance scale is assumed.

{}From optical observations the stellar iron abundances in early-type 
galaxies are believed to range from $\sim 1/3$ solar for dwarf
ellipticals such as M32 (Freedman 1989) to a few times solar for 
giant ellipticals (e.g., O'Connell 1986; Bica 1988).
The iron enrichment due to SNIa's is critically sensitive to the adopted
SN rate in ellipticals, a quantity that is still affected by a major
uncertainty. van den Bergh \& Tammann (1991) report a high value
($\vartheta_{\rm SN}=1.1$) virtually identical to the old estimate of Tammann
(1982). On the other hand, Cappellaro et al. (1993) estimate a substantially
lower value ($\vartheta_{\rm SN}=0.25$). Thus, combining into equation 
(1) the estimated stellar  abundances with the SN contribution, we 
estimate the expected iron abundance in the hot ISM of elliptical 
galaxies to range from a minimum of $\sim 2$ times solar, to perhaps 
as much as $\sim 5$ times solar or more, depending on the adopted SNIa rate.

The determination of the iron abundance of the hot ISM of elliptical
galaxies offers the opportunity to check at once the average stellar
metallicity as estimated from optical observations, and the expected
contribution from SNIa's. This latter aspect is particularly important as this
SN rate  is one of the key parameters controlling the gas flow regime
that is established in hydrodynamical models of elliptical galaxies
(Ciotti et al. 1991;
Renzini 1994a). As extensively discussed by these authors, crucial for the
evolution of the gas flows over cosmological times is also the secular
evolution of the SNIa heating after the completion of the bulk of star
formation in these galaxies. For exploratory purposes 
such heating can be assumed to decline with 
time as a power law, i.e., $\propto t^{-s}$, a convenient
parameterization. In absence of other energy sources, the gas flows evolve
from early supersonic winds, to subsonic outflows, and eventually to inflows
if $s\gsim 1.3$. Conversely, the reverse sequence of flow regimes is established
if $s\lsim 1.3$, with early inflows turning to outflows and winds at later
times (Loewenstein \& Mathews 1991; David, Forman, \& Jones 1990).
The epoch of transition from one flow regime to another is very sensitive
to several galaxy parameters, especially to its mass, luminosity, structure,
and SN rate. As both SN heating and iron enrichment of the ISM depend on the
assumed SN rate, it was soon realized that the determination of 
the iron abundance in the ISM could lead to a welcome reduction in 
the number of parameters, thus setting an important constraint on the 
evolution of gas flows in galaxies.

An estimate of the iron abundance in elliptical 
galaxy flows was first attempted using data
from the {\it Ginga} satellite. Ohashi et al. (1990) reported an upper limit
about twice solar for NGC 4472 and NGC 4636. From the upper limit to the iron
line equivalent width reported by Awaki et al. (1991) for NGC 4472, Ohashi \&
Tsuru (1992) estimate the iron abundance in the flow to be at most solar.
For NGC 1399 -- the cD galaxy in the Fornax cluster -- Ikebe et al. (1992)
estimate iron to be $1.1^{+1.3}_{-0.5}$ times solar. From {\it BBXRT}
observations and one-temperature fits Serlemitsos et al. (1993) estimate the
metallicity of NGC 1399 and NGC 4472 to be at 90\% confidence $0.56^{+0.82}_
{-0.38}$ and $0.20^{+0.46}_{-0.09}$, respectively, while from {\it ROSAT}
observations Forman et al. (1993) infer iron to be 1--2 times solar 
in this latter galaxy.

Iron abundance determinations in elliptical galaxy flows has then gained
great impetus by the advent of the {\it ASCA} satellite, thanks to its
superior spectral energy resolution. The low iron abundances appear to be
confirmed by {\it ASCA} observations (Awaki et al. 1994; see also \S 2.2),
that indicate e.g., for NGC 4472 an iron abundance $0.52^{+0.09}
_{-0.07}$ at the 90\% confidence level.

There is clearly a macroscopic discrepancy between the expected
abundance, even with the lowest SNIa enrichment, and what is consistently 
indicated by the X-ray observations of elliptical galaxies with four 
different X-ray telescopes: {\it Ginga, BBXRT, ROSAT}, and {\it ASCA}. 
Instrumental problems, such as calibrations and the like, can
therefore be firmly excluded as the origin of this {\it iron  discrepancy.}
Note also that the discrepancy is exacerbated by another factor $\sim 2$ if
 a short distance scale is adopted ($h_{50}=2$).
We believe that the solution of the discrepancy, whatever it is, will
have profound implications for either our understanding of galaxy formation
and evolution and/or for the X-ray diagnostics of astronomical objects, and in
this paper we thoroughly address the question. In \S 2 we present and discuss
the state of the art estimates for the iron content of the stellar and diffused
components of elliptical galaxies, as derived respectively from integrated
optical spectra and X-ray observations.
In \S 3 we discuss several astrophysical implications of the observed
low iron abundances in the ISM, assuming these determinations to give the
actual abundance to be compared with the prediction of equation (1). These
implications include the role of SNIa's in ellipticals, the chemical
evolution of these galaxies, as well as the the enrichment and heating of the
intracluster medium. Having encountered several astrophysical difficulties
that are generated by this assumptions, in \S 4 we explore various alternative
possibilities of hiding or diluting iron in the ISM, so as to reconcile
the expectations with the X-ray observations. These possibilities include
dilution of the ISM with iron-poor gas from an intracluster medium (ICM),
as well as astration from the ISM, either in the form of particulates
(iron flakes), or of substellar mass objects (Jupiters).
None of the explored solutions appears very attractive or astrophysically
plausible, and in \S 5 we alternatively rise the question of the reliability
of iron line diagnostic tools that are currently used in conjuction with
X-ray observations. At the temperatures that
are typical of the hot ISM of elliptical galaxies ($\lsim 1$ keV), 
the iron abundance is derived from the strength of several lines 
originated by electron transitions down to the L-shell
in incompletely ionized iron ions, typically Fe XVIII-XXI, hence
called Iron-L. Conversely, X-ray Iron-K lines are due to electron decays to
the K shell of He-like and H-like iron ions. In \S 5 we compare 
the results of Iron-L diagnostics at different temperatures according
to different thin plasma models, and finally we list and discuss the iron 
abundance in a variety of astrophysical objects as derived from {\it ASCA} 
data and current Iron-L diagnostics. Our conclusions are presented in \S 6.

\medskip
\centerline{\bf 2. THE IRON CONTENT OF ELLIPTICAL GALAXIES}
\medskip
\centerline{\it 2.1. The Iron Abundance in the Stellar Component}
\medskip

As already mentioned, the stellar iron abundances in early type 
galaxies are believed to range from $\sim$ 1/3 solar in dwarf
ellipticals such as M32 to a few times solar in giant elliptical galaxies. 
{}From a theoretical work, which itself is an extended version of Larson's
(1974) supernova-driven galactic wind model, Arimoto \& 
Yoshii (1987) first obtained a stellar metallicity distribution 
ranging from $\sim$ 1/10 solar to about 10 times solar for the bulk 
of stars in a giant elliptical galaxy, with a luminosity-weighted mean 
$\sim$ 2 solar. Such a huge amount of heavy elements is
produced during an intensive phase of star formation in the early stage
of galaxy formation. As the broad-band colors of ellipticals are well
matched by Arimoto \& Yoshii models, it has been widely accepted that  giant 
elliptical galaxies are super metal-rich, at least $\sim$ 2 solar or even more.

In this section the iron abundance is now derived from the observed
Mg$_2$ indices by using the relation given by the population synthesis 
model of Buzzoni, Gariboldi, \& Mantegzza (1992):

     $$\feh = 7.41 Mg_2 -2.07. \eqno (2)$$
The population synthesis models of Worthey (1994) give a slightly different 
calibration, i.e., $\feh=5.85$Mg$_2-1.66$, but the resulting values of 
$\feh$ are almost identical to those calculated with equation (2). 
We note that both  relations give considerablly steeper
slopes than the early population synthesis models by Mould (1978), for which 
$\feh=3.9$Mg$_2-0.9$. 

Giant elliptical galaxies usually exhibit conspicuous color and 
metallic line-strength gradients (e.g., Franx, Illingworth, \& Heckman
1989; Peletier et al. 1990; Gorgas, Efstathiou, 
\& Arag\'on Salamanca 1990; Davies, Sadler, \& Peletier 1993).
This implies a negative radial gradient of the stellar metallicity,
and therefore the true {\it mean} iron content 
of an elliptical galaxy may be appreciably  lower than the value indicated by
colors or spectral features taken at the galaxy center.

The line-strength gradients in elliptical galaxies are considered to
provide clues to the importance of dissipative processes in galaxy
formation (Larson 1976; Carlberg 1985). In an early study, Faber (1977)
first measured radial gradient in the Mg$_2$ index for NGC4472 and 
reported that the contour of constant line-strength tends to be flatter 
than the isophote, in agreement with Larson's dissipative collapse models. 
In recent years, the line-strength gradients have attracted much attention
(e.g., Efstathiou \& Gorgas 1985;  Baum, Thomsen, \& Morgan 1986;
Thomsen \& Baum 1987; Couture \& Hardy 1988; Gorgas et al. 1990;
Boroson \& Thompson 1991; Delisle \& Hardy 1992; Davies et al. 1993;
Carollo \& Danziger 1994a,b) bringing above $\sim 40$ the number  of 
elliptical galaxies with measured line-strength gradients. 
In this section, we present a general method to estimate the  
mean iron abundance of stars in an elliptical galaxy that takes 
gradients into account, and give the resulting iron contents for 
$\sim$ 40 galaxies with known line-strength gradients.

It is widely known that the surface brightness distributions of elliptical 
galaxies are well fitted by the so-called $r^{1/4}$-law (de Vaucouleurs 1948):

     $$I(r)=\ie\exp \lbrace -b [({r \over \re})^{1/4}-1] \rbrace, \eqno (3)$$
where $b=3.33 \ln 10$ and the length scale $r_e$ is  the effective 
radius, interior to which one-half of the total light of the galaxy is 
emitted. Thus, the total luminosity (e.g., $\lb$) is given by:

$$\lb=\int_0^{\infty}2\pi rI(r)dr = 8\pi r_{\rm e}^2\ie e^b\Gamma(8)b^{-8},
     \eqno (4)$$
where $\Gamma$ is the gamma function.

For the radial distributions of the iron abundance $Z^{\rm Fe}(r)$ 
we adopt the convenient parameterization: 

     $$Z^{\rm Fe}(r)=Z^{\rm Fe}(\re)({r \over \re})^{-c}, \eqno (5)$$
where both $Z^{\rm Fe}(\re)$ and the {\it slope} parameter $c$ can be 
derived from the line-strength measurements after properly transforming the 
line index to the iron abundance. Usually the line-strength measurements do
not reach much  beyond the effective radius, but here we
assume that equation (5) holds for the whole  galaxy, a reasonable
approximation in most cases. Indeed, only few ellipticals show steeper
line-strength or color gradients beyond $\re$, such exceptions being 
e.g., NGC 5813 (Gorgas et al. 1990), NGC 4697
(Franx et al. 1989), and NGC 4889 (Peletier et al. 1990).
The total iron mass locked in the stars is then given by:

     $$M^{\rm Fe}=({M_* \over \lb})\int_0^{\infty}2\pi 
      Z^{\rm Fe}(\re) r I(r) ({r \over \re})^{-c}dr=$$
     $$=8\pi r_e^2I_e({M_* \over \lb}) 
      Z^{\rm Fe}(r_e)e^b\Gamma(8-4c) b^{-(8-4c)}=$$
     $$=M_*{\Gamma(8-4c) \over \Gamma(8)}b^{4c}Z^{\rm Fe}(\re) ,
     \eqno (6)$$
where $M_*$ is the total stellar mass. In deriving equation (6) 
the stellar mass-to-light ratio is assumed to be constant throughout a galaxy.
The mean iron abundance of stars is then given by:
     $$<Z^{\rm Fe}_*> \equiv {M^{\rm Fe} \over M_*}
= {\Gamma (8-4c) \over \Gamma (8)} b^{4c}
Z^{\rm Fe}(\re)=$$
$$=\beta(c) Z^{\rm Fe}(r_e), \eqno (7)$$
where $\beta(c) \equiv b^{4c}\Gamma (8-4c)/\Gamma(8)$.
Equation (7) can also be written as:
     $$<\!\feh\!> = \feh _{r_e} + \log \beta(c), \eqno (8)$$
and the function $\beta(c)$ is, with a few exceptions, close to
 unity for the observed range of $c$ values, as shown in Fig. 1. 
Thus,  {\it the mean, luminosity-weighted iron 
abundance of stars in an elliptical galaxy is to a good approximation
given by a value 
measured at the effective radius $r_e$, or $<\!\feh\!> \simeq \feh_{\re}$}.
A correction to $\feh_{\re}$ is at most +0.3 dex, but usually
less than +0.2 dex.

Note that the observed Mg$_2$ gradients coupled with the calibrations by 
Buzzoni et al. (1992) and Worthey (1994) give an average iron abundance 
gradient $\Delta \feh/$ $
 \Delta \log (r)  = -0.5 \pm 0.2$, or a reduction by 50 
to 80\% over a factor of 10 in radius:
A significant reduction that is consistent with the dissipative
models (Larson 1974; Carlberg 1985).

Table 2 gives the resulting mean iron abundances of stars in the
sample of elliptical galaxies; column (2) gives $\re$ in arcsec with 
the corresponding reference in column (3); columns (4) and (5) give 
the Mg$_2$ index at galaxy center and at $r=\re$, respectively, with the 
corresponding source in column (7); column (6) gives the Mg$_2$ gradient
defined as $d \equiv \Delta Mg_2/\Delta \log (r)$. A value of $d$
is estimated for each galaxy by applying a linear least square fit
to the observed Mg$_2$ values except for the most inner region with
$\log (r/r_e) \lsim -1.0$. Columns (8) and (9) give central and mean
iron abundances, respectively, the latter from 
equation (8); finally, column
(10) gives the stellar velocity dispersion at galaxy center
from Davies et al. (1987). The iron abundance {\it slope} is then 
given as $c=-7.41d$, where we have used equation (2).
It appears that in galaxies with $\sigma_0 \gsim 250$ km s$^{-1}$
the average {\it central} iron abundance is
in the range [Fe/H]$_0 \simeq 0.24 \pm 0.10$, while the {\it mean} iron 
abundance is $<\!\feh\!> \simeq -0.06 \pm 0.13$, i.e., a factor $\sim
2$ lower ($\sim$ solar).

In essence these values are luminosity-averaged abundances, the
weighting luminosity being that of the continuum in the vicinity of
the Mg$_2$ feature (i.e., about $\lv$-averaged). Correspondingly, they
should be regarded as lower limits to the {\it 
mass-averaged} iron abundance, as the most metal rich components in
the distribution of metallicity have lower $\lv$ luminosity for given
mass $M_*$ (Greggio 1996b). The size of the effect is very
sensitive to the detailed metallicity distribution, and for the
closed-box model the 
mass-averaged abundance can be $\sim 2$ times larger than the 
luminosity-averaged abundance. Hence, the values of $<\!\feh\!>$ in
Table 1 should be regarded as lower limits.

\centerline{\it 2.2. The Iron Abundance in the Hot ISM}
\medskip
With its two detectors (SIS and GIS), the {\it ASCA} X-ray astronomy
satelite  is capable of performing
imaging and spectroscopic observations with high energy resolution, 
simultaneously over the energy range from 0.4 keV to 10 keV 
(Tanaka, Inoue, \& Holt 1994; Ohashi et al. 1996; Makishima et al.
1996). A number of emission lines from  elements such as oxygen,
silicon, sulfur, and iron can be identified, and the 
intensity of these lines allow the determination of the 
abundances in the hot ISM, and iron abundances for several elliptical
galaxies have already been reported (Awaki et al. 1994; Loewerstein et al.
1994; Mushotzky et al. 1994). 

     {\it ASCA} has also directly confirmed 
the presence of a hard X-ray component with
$kT > 2$ keV (Matsushita et al. 1994) which was suggested by 
{\it Einstein} and {\it ROSAT} observations of low $L_X/L_B$ elliptical 
galaxies (Kim, Fabbiano, \& Trinchieri 1992; 
Fabbiano, Kim, \& Trinchieri 1994). The luminosity of the hard 
component of the observed galaxies  is similar to the contribution by
discrete sources as estimated by 
Canizares, Fabbiano, \& Trinchieri (1987) for 
spiral galaxies. That is, the number of low mass X-ray binaries
(LMXRBs) per unit 
blue-luminosity that emit the hard component appears to be the same in
ellipticals as in spirals.
 We therefore assume that the hard component due to the discrete
sources  is harder than 5 keV when fitted with bremsstrahlung
spectrum. It is worth noting  that the lower limit to the abundance is
not so sensitive to the adopted temperature of the hard 
components for $kT_{\rm hard} \gsim $ 2 keV, but the
upper limit can instead become quite large.

{\it ASCA} data are now available for several elliptical galaxies,
and  for a subsample of them (NGC720, NGC1399, NGC1404, NGC4374,
NGC4406, NGC4472, and NGC 4636) 
we have reanalyzed the spectra and determined the iron 
abundances with several different plasma models (see also section 5.1). 
These are all bright ellipticals, with
$\lb$ ranging  from $\sim 10^{10}$ to $10^{11}\lsun$, and fairly high
$\lx/\lb$ ratio.
For low $\lx/\lb$ galaxies the hard component dominates the spectra 
and it is difficult to determine the ISM abundance; thus these galaxies 
are not considered in this paper. 
The observed galaxies are located in a variety of physical  enviroments:
NGC4406, NGC4374, NGC4472, and NGC4636 are in the Virgo Cluster.
{\it Ginga} detected the dense ICM around these galaxies except 
NGC4636 (Takano 1989).
NGC4636 is far from Virgo Cluster center and 
is surrounded by very extended X-ray emission characteristics of
group (Trinchieri et al. 1994).
NGC1399 is the cD galaxy in the Fornax cluster, and NGC1404 is located
near NGC1399. 
{\it ASCA} detects the ICM of the cluster around these galaxies
(Ikebe 1996). Therefore, NGC720 is the only truely isolated system 
and others may be within the ICM.

   We have integrated the SIS and the GIS spectral data within a
radius of $3'$
for low $\lx/\lb$ galaxies (Log $\lx/\lb <$ 30.7), and within $5'$ for high
$\lx/\lb$ galaxies. 
The background was removed by subtracting a spectrum of a 
blank sky emission of the same area using the same part of the detector.
For NGC720, NGC1399, NGC4472, and NGC4636, the background has been removed
by subtracting a spectrum of a blank sky emission of the same area using 
the same part of the detector. For NGC1404, NGC4374, and NGC4406, there
is emission from the surrounding ICM, so that we also subtract the surrounding 
ICM as background; thus we accumulate spectra over the region of the same
radius whose angular distance to the optical axis of the XRT is nearly the 
same as the background spectra. For NGC1404, the region accumulated as 
the background is nearly at the same angular distance to the cD galaxy, 
NGC1399. The spectra have then
been fitted with both $\chi^2$
and maximum likelihood methods using the XSPEC spectral fitting
package (Raymond \& Smith 1977).
Both methods give nearly identical results, so we discuss only the results
with the $\chi^2$ method. Thin thermal plasma models 
modified by interstellar absorption with solar abundance ratios are
fitted to each galaxy, and a 
bremsstrahlung spectrum is  assumed for the hard component. 
The iron abundances are determined from the iron-L blends around 1 keV,
and the fitting results of 90\% confidence
are given in Table 2, assuming the solar iron 
abundance to be $4.68\ 10^{-5}$ by number. The derived temperatures of
the ISM are in the 
 range from $\sim$0.6 keV to $\sim$1.1 keV, while the iron
abundances  range  from $\sim$0.1 
solar to $\sim 0.4$ solar. XSPEC further returns
hydrogen column densities  
$\sim 10^{21}\rm{cm^{-2}}$,  significantly higher than Galactic values 
$\sim 1-3\ 10^{20}\rm{cm^{-2}}$.
The temperature and the iron abundances obtained
here are consistent with the results by {\it Ginga} and {\it BBXRT} 
observations (Awaki et al. 1991; Ikebe et al. 1992; Serlemitos et al. 1993). 
We note that the fit parameters  are
consistent with {\it ROSAT} 
results for NGC4636 (Trinchieri et al. 1994), but 
disagree with the {\it ROSAT} observation for NGC4472 (Forman et al. 1993).
The iron abundance obtained by {\it ROSAT} for NGC4472 is consistent
with being  solar, 
while the hydrogen column density is consistent with the Galactic value.
These abundances derived here are systematically different by factor
1.5 from those by Awaki 
et al. (1994), who adopted a  different solar iron
abundance. The iron abundances listed in Table 2 are consistent with
the results obtained by Loewenstein et al. (1994), Mushotzky et al.
(1994), and Matsushita et al. (1994), though Loewenstein et al. did not
include the hard component in their spectrum fitting. When the contribution
of the ICM is subtracted, the fluxes of hard component of NGC1404 and NGC4374 
are consistent with other galaxies when scaled with their optical
luminosities (cf., Matsushita et al. 1994).

     Figures 2 and 3 show the resulting iron abundances as a function of
$L_X/L_B$ and temperature, respectively. There is a marginal trend of 
the iron abundance with both temperature and the X-ray to
blue-luminosity ratio.  The four galaxies with highest temperature or
$\log L_X/L_B$ ratio appear to have
 iron abundances $\sim 0.3-0.4$ solar, while the three galaxies 
with lower $\lx/\lb$ ratio  abundances are even smaller, $\sim 0.1-0.2$
solar. Among galaxies listed in Table 1 four  are 
common with our {\it ASCA}
analysis (namely, NGC4374, NGC4406, NGC4472, and NGC4636).
The derived ISM iron abundances of NGC 4374, NGC 4406, 
NGC 4472 and NGC 4636 are
0.1, 0.2, 0.3, and 0.3 solar, respectively.
In contrast, the mean stellar iron abundances of these galaxies are 0.9,
0.7, 1.4, and 0.7 solar, respectively. NGC720,
NGC1399, and NGC1404 are not listed in Table 1, but judging from the
velocity dispesion of these galaxies  ($\sigma_m \simeq 230 - 310$
km s$^{-1}$) one can expect their mean stellar iron abundances 
to be also $\sim$ solar.
In addition, according to equation (1), 
the SNIa contribution to the ISM  iron is at least 1$\times$solar,
when the lowest SNIa rate is assumed 
(Cappellaro et al. 1993). The iron abundance of ISM should then be at least 
twice solar, and the discrepancy factor appears to range from $\sim
4$ to $\sim 20$. It may actually be even larger, considering that
stellar abundances in Fig. 2 are luminosity-, rather than mass-weighted.

     We note that the silicate discrepancy, if any, is not as significant
  as the iron. The silicate abundance can be determined from the
  silicate K-line, and the atomic physics of this He-like Si line are 
  believed to be much secure than the iron-L line, although one needs to
  evaluate precisely the hard component in this energy range. We have 
  estimated the silicate abundance for X-ray luminous ellipticals, NGC1399,
  NGC4406, NGC4472, and NGC4636, by using the higher energy parts 
  ($kT \ge 1.7$ keV) of the same spectra. Raymond-Smith model with the
  solar abundance ratios is fitted to each galaxy, and a 10 keV bremsstrahlung 
  spectrum is assumed for the hard component. The resulting temperatures 
  are consistent with those given in Table 2 for NGC4406 ($0.86^{+0.17}_
  {-0.11}$ keV) and NGC4636 ($0.78^{+0.10}_{-0.06}$ keV), but slightly 
  higher for NGC1399 ($1.27^{+0.11}_{-0.11}$ keV) and NGC4472 
  ($1.05^{+0.06}_{-0.06}$ keV). The silicate abundance is in average half
  solar; or more precisely, $0.43^{+0.20}_{-0.13}$ solar for NGC1399, 
  $0.53^{+0.48}_{-0.37}$ solar for NGC4406, $0.48^{+0.26}_{-0.17}$ solar 
  for NGC4472, and $0.60^{+1.00}_{-0.17}$ solar for NGC4636. 
  Recently, NGC4636 is re-observed during {\it ASCA} AO4 run with an
  exceptionally long exposure time (200 ks) by T. Ohashi et al. (1996, private
  communication), and the silicate abundance for this galaxy is now
  believed to be $1.0^{+0.3}_{-0.3}$ solar. Since the bulk of silicate 
  is synthesized in SNII and comparatively less in SNIa, the X-ray 
  abundance of silicate should be nearly equal to the stellar silicate
  abundance. If the stars have the same silicate abundance as the iron, 
  this should be about solar; thus both the optical and the X-ray silicate
  abundance is consistent with each other for NGC4636. For other three
  galaxies, there is no reason for not expecting a similar increase in
  the measured silicate abundance, once much higher signal-to-noise ratios
  are achieved in their spectra.

\medskip
\centerline{\bf 3. ASTROPHYSICAL IMPICATIONS OF A LOW IRON}
\centerline{\bf IN THE HOT ISM}
\medskip

A comparison of the iron abundances as inferred from optical observations
of the starlight and from X-ray observations of the hot ISM has then revealed
a major discrepancy: even neglecting any ISM enrichment from SNIa's, the 
stellar
iron abundances exceed those derived for the hot ISM by factors that range
between $\sim 2$ and $\sim 10$ (cf. Table 1 and 2, and Fig. 2).
We believe that the {\it optical} and the X-ray
abundances cannot be easily reconciled, and therefore the existence of this
macroscopic discrepancy opens three main options: 1) the optical abundances
are seriously in error {\it and} the SNIa iron enrichment is much
lower than
currently believed, 2) there is a problem with the interpretation of the
Iron-L lines, and 3) both optical and X-ray abundances are correct,
but the ISM
iron is somehow hidden to X-ray observations.
In this section we assume 
the ISM iron abundances inferred from the X-ray observations and reported in \S 2.2 to be correct, 
and explore the astrophysical consequences that stem from such
assumption.

The low iron abundances reported for some galaxies (especially NGC 1404 and
NGC 4374) require the SNIa iron enrichment at the present time to be vanishingly
small, and the average iron abundances of the stellar component to be in error
by a large factor. The first requirement has been extensively discussed
in RCDP. Even with a SNIa rate as low as advocated by Cappellaro et al. (1993),
$\vartheta=1/4$, SNIa's would enrich the ISM to well above solar if each SNIa
releases $0.7\msun$ of iron. Thus, the implications of option 1) for SNIa's
include a drastic reduction of the SNIa rate in ellipticals over the most
conservative estimates, and/or a drastic reduction of the iron yield per SNIa
over current theoretical estimates. In practice, X-ray observations imply
that the product $\vtsn(M^{\rm Fe}/0.7\msun)$ 
should be much less than $\sim 1/4$,
thus challenging at once both the theory and the observations of SNIa's.

Concerning the stellar abundances, we note that the existing $\mg2$-[Fe/H]
calibrations are far from being exempt from uncertainties, as they are
based on elaborate synthetic population modelling and might
be in error for a variety of reasons. Moreover,  the $\mg2$ index is
a rather indirect measure of iron abundance, while the [Mg/Fe] ratio may not
be solar in the stellar component of elliptical galaxies, with
Mg being enhanced compared to Fe by perhaps as much as a factor of $\sim 2$
(e.g., Worthey, Faber, \& Gonz\'alez 1992; Davies, Sadler, \& Peletier 1993).
If so, the discrepancy could be somewhat alleviated, but clearly not removed.
Again, a combination of an erroneous $\mg2$-[Fe/H] calibration and an
underestimate of the metallicity gradients within galaxies should be required
to explain the much larger iron abundances inferred from optical observations,
compared to those derived from the X-ray data. We believe this to be a very
drastic requirement, and  emphasize that it would have a major
impact on the stellar population studies of elliptical galaxies, and therefore
on our understanding of their formation and evolution.

\medskip
\centerline{\it 3.1. Type Ia Supernovae and Gas Flows in Elliptical Galaxies}
\medskip

The drastic reduction in the rate of SNIa's demanded by the X-ray data has
major implications for the interpretation of the X-ray properties and evolution
of elliptical galaxies. With $\vtsn\ll 1/4$ the SN heating of the ISM becomes
virtually negligible for the dynamics of the gas flows in these galaxies at the
present time. In absence of alternative internal sources of heat, 
the ISM of virtually all elliptical galaxies would
now be in a cooling flow regime. Correspondingly, all elliptical
galaxies should
be very bright in X-rays, at variance with the observations that indicate a
large disperion in X-ray luminosity for given optical luminosity 
(Canizares et al. 1987; Donnelly, Faber, \& O'Connell 1990;
Ciotti et al. 1991; Fabbiano, Kim, \& Trinchieri 1992; Renzini 1996b). 
With sufficient SN
heating instead, most galaxies are in a wind or outflow regime, with
substantially lower X-ray luminosity compared to inflow models.
To account for the X-ray luminosity dispersion even in absence of SN heating,
an alternative
heating source seems to be required so as to maintain the majority of
ellipticals in a wind or outflow regime at the present time. This may be
represented by sporadic AGN activity  triggered each time that
a central cooling catastrophe turns the gas flow from an outflow to
an inflow
regime, thus fueling
a previously starving central black hole (Ciotti et al. 1991; Binney
\& Tabor 1995; Ciotti \& Ostriker 1996; see also
Renzini 1994a, 1996b). Alternatively, it has been suggested that
ram pressure stripping may be responsible for the partial or total removal of
the hot ISM in some galaxies, thus
reducing the amount of X-ray emitting gas and X-ray
luminosity by appealing to environmental effects
(White \& Sarazin 1991). This explanation
would predict elliptical galaxies in the
Coma cluster to be on average fainter in X-rays than ellipticals in Virgo of
similar optical luminosity, a result of the somewhat higher ram pressure in
Coma compared to Virgo. However, {\it ROSAT} observations of individual
galaxies in Coma do not reveal systematic differences with respect to those in
Virgo (Dow \& White 1995), which argues against strong environmental effects
being responsible for the X-ray luminosity dispersion among galaxies in these
two clusters. 

In conclusion, if one excises SNIa's from the factors affecting gas flows in
today ellipticals, alternative mechanisms able of driving gas out of galaxies
have to take their place on the stage. A prime candidate is sporadic (or
recurrent) AGN
activity, though ram pressure stripping cannot be fully excluded at the
present time.

\medskip
\centerline{\it 3.2. SNIa Rate in the Past} 
\medskip

Over the last 15 years the notion of a prime role of SNIa's in the
production of iron in the Galaxy has become widely accepted.
The basic fact is represented 
by the O/Fe and $\alpha$-elements over iron ratios
in metal poor stars of the Galactic Halo, that is 3-4 times solar 
(Sneden, Lambert, \& Whitaker
1979; Gratton \& Ortolani 1986; Bessell, Sutherland, \& Ruan 1991; 
Barbuy 1992; see also the review by Wheeler, Sneden, \& Truran, 1989).
The interpretation of the observed trend of [$\alpha$/Fe] vs [Fe/H]
is based
on the different nucleosynthesis yields and timescales of SNIa's and SNII's,
with the
enrichment of the Halo being sufficiently rapid to sample almost exclusively
the
prompt release of SNII products (primarily the $\alpha$-elements such as
O, Ne, Mg, Si, etc., but comparatively little iron), while it takes a
longer time for SNIa's to release the bulk of
their iron (e.g., Sneden et al. 1979; Greggio \& Renzini 1983a,b; Matteucci \&
Greggio 1986; Abia, Canal, \& Isern 1991; Matteucci \&
Fran\c cois 1992; Timmes, Woosley, \& Weaver 1995; Greggio 1996a;
Ruiz-Lapuente, Burkert, \& Canal 1996). 
In this scenario, to which we refer
as the {\it Standard Chemical Model} (SCM),
SNII's had produced virtually all the iron now locked in extreme population II
stars, while at later times SNIa's would have taken the lead as main iron
producers, to the extent that $\sim 3/4$ of the iron present in the sun would
come from SNIa's (Timmes et al. prefer $\sim 1/2$). Additional support to the SCM comes from star forming
(spiral) galaxies exhibiting a SNIa rate about 1/3 that of all other types
together (Cappellaro et al. 1993). If each SNIa produces $\sim 10$
times more iron than other SN's as currently believed, then the present iron
production rate in spiral
galaxies is indeed dominated $\sim 3$ to 1 by SNIa's (see RCDP).

The SCM has a series of testable predictions when applied to elliptical
galaxies and clusters of galaxies, if one assumes that IMF and SNIa
productivity are the same as in our own Galaxy. The first prediction is that
the past time-average SNIa rate in ellipticals had to be much higher than the
present rate, otherwise SNIa's would have produced a negligible
fraction of all the iron that is observed in clusters of galaxies (Ciotti et al.
1991; Arnaud et al. 1992; RCDP). More quantitatively, with $\vtsn=1$ 
such average rate should be
$\sim 10$ times higher than the present rate in ellipticals in order for
SNIa's to have produced $\sim 3/4$ of the iron that is observed in
clusters. 
A reduction
of the present rate by e.g., a factor of 4 (i.e., to $\vtsn=1/4$) does not
necessarily
invalidate this scenario: suffice to increase by the same factor the past
average rate. In the frame of the parameterization mentioned in \S 1, this
would require  increasing the slope $s$ from $\sim 1.4$ to $\sim 1.7$.
This is not at all inconceivable, given our enduring ignorance of
the nature of SNIa progenitors (Renzini 1996a; Branch et al. 1995). 

In this context, it is worth emphasizing that the present iron
abundance in the ISM of ellipticals can effectively constrain only
the recent SNIa rate, as -- with the exception of a few extreme cases
(Loewenstein \& Mathews 1991) -- the flow time of the gas in or out
the galaxy is just a few Gyr (Ciotti
et al. 1991). In particular, no constraint is placed on first $\sim 10$ Gyr,
when indeed SNIa's would have delivered most of their iron yield.
We conclude that a moderately low iron abundance in the ISM of today
ellipticals is certainly a challenge, 
but does not necessarily invalidate the
applicability of the SCM also to elliptical galaxies.

\medskip
\centerline{\it 3.3. Enrichment of the Intracluster Medium}
\medskip
\centerline{\it 3.3.1 Iron Content of Galaxy Clusters}
\medskip
In this section we explore further consequences of a low iron content of
elliptical galaxies that may be relevant for  our understanding of
the chemical and thermal evolution of the ICM.

A very effective way of quantifying the iron content of clusters is  by their
iron mass to light ratio (IMLR), i.e., by the ratio of the total mass of iron
in a cluster
over the total optical luminosity of the cluster galaxies. The IMLR can be
defined
separately for the ICM, for the galaxies themselves, as well as for the
cluster as a whole.
The IMLR for the ICM of clusters appears to be remarkably constant,
irrespective of the cluster optical luminosity (RCDP; Renzini 1994b,
see its Fig. 1):
$$\left({M^{\rm Fe}\over\lb}\right)^{\rm ICM}\simeq (0.01-0.02)h_{50}^{-{1\over
2}}\quad\quad (\msun/\lsun),\eqno(9)$$
The IMLR for the iron now locked into stars inside cluster 
galaxies is given by:
$${M^{\rm Fe}_*\over\lb}=<\! Z^{\rm Fe}_*\!>{M_*\over\lb}\simeq
(0.01-0.02){<\! Z^{\rm Fe}_*\!>\over Z^{\rm Fe}_{\odot}} h_{50}\quad
      (\msun/\lsun), \eqno(10)$$
where $<\! Z^{\rm Fe}_*\!>$ is the average iron mass fraction in the
stellar component of cluster galaxies, and $M_*/\lb$ is the stellar mass to
light ratio of galaxies. A comparison between equation (9) and equation (10)
reveals that -- assuming the average stellar iron to be solar -- there is
a nearly equal amout of iron in the ICM as there is locked into stars (for
$h_{50}=1$). This {\it fifty-fifty} condition has been extensively discussed.
Note that the total IMLR is not very sensitive to the adopted
value of the Hubble constant, but the relative share of iron between the two
cluster components shows a significant dependence (Renzini 1994b). For
example, with $h_{50}=2$ the fifty-fifty share turns to a 3 to 1 share in favor
of galaxies, i.e., the ICM holds $\sim 1/4$ of the total iron, cluster galaxies
3/4.
The  iron share between the two cluster components sets a strong
constraint on models of galaxy formation and evolution: nearly as much iron
needs to be ejected from galaxies as remains locked into stars.

These figures assume $<\! Z^{\rm Fe}_*\!>\simeq Z^{\rm Fe}_{\odot}$. Now,
if the abundances derived from the X-ray data are correct, the average
metallicity of stars is appreciably below solar. To illustrate the impact of
the new abundances, we take for $<\! Z^{\rm Fe}_*\!>$ the straight average
of the values in Table 2, i.e., $\sim 0.2$ times solar. With this value
equation (10) gives a stellar IMLR $\sim 1/5$ that of the ICM, i.e., $\sim 5/6$
of the total iron is now in the ICM and only 1/6 remains in stars. 
Thus, the iron abundances from the X-ray data  imply a much more
dramatic degassing of 
 galaxies, that should have lost perhaps more than 80\% of their initial mass.
Such rather extreme consequences can be partly alleviated by
appealing to a shorter extragalactic distance scale: with $h_{50} =2$ the iron
share becomes $\sim 2$ to 1 in favor of the ICM, with galaxies having lost
$\sim 60\%$ of their initial mass. 
In conclusion, the low
iron abundances in the ISM of ellipticals that are indicated by the X-ray
data have profound implications for the formation process of galaxies,
dramatically altering  the mass budget, i.e., the ratio of the (baryonic)
mass ejected by galaxies in the course of their formation and evolution,
to the residual galaxy mass (in stars).

This argument can be followed a little further. We take the Coma cluster as
a representative rich cluster. The total mass of the ICM is estimated to be
$\sim 3.1\times 10^{14}h_{50}^{-5/2}\,\msun$ and the total stellar mass in
galaxies $\sim 2.0\times 10^{13}h_{50}^{-1}\,\msun$ (White et al. 1993).
Thus the ICM to star mass ratio is $\sim 15$ to 1 for $h_{50}=1$, or $\sim 5$
to 1 for $h_{50}=2$. Thus, the fraction of the ICM that comes from galaxies
is about 1/3, irrespective of $h_{50}$, with the residual 2/3 being represented
by pristine material, never incorporated into galaxies. 
In the traditional view
instead, with $<\! Z^{\rm Fe}_*\!>\simeq Z^{\rm Fe}_{\odot}$, only a much
smaller fraction of the ICM, i.e., $\sim 1/15$, was once inorporated into
galaxies.
\medskip
\centerline{\it 3.3.2 Heating of the ICM}
\medskip
The presence of a large amount of iron in the ICM implies that a large
amount of matter has been lost by galaxies. Arguments are presented in RCDP
supporting the notion that matter was  ejected rather than swept  by ram
pressure stripping, as also indicated by the IMLR of the ICM being independent
of cluster richness. Thus, along with mass, galaxies have also injected
kinetic energy (i.e., heat) into the ICM.
The kinetic energy associated to galactic winds, again per unit cluster light,
is given by 1/2 the ejected
mass times the square of the typical wind velocity (e.g. Renzini 1994b), i.e.:
$${E_{\rm w}\over\lb}={1\over 2} \,\left({M^{\rm Fe}\over\lb}\right)^{\rm ICM}
\left<\!{v_{\rm w}^2\over
   Z^{\rm Fe}_{\rm w}}\!\right>\simeq 1.5\times 10^{49}{1\over Z^{\rm Fe}_{\rm
   w}/Z^{\rm Fe}_\odot}\cdot
   \left({v_{\rm w}\over 500\,{\rm km}\,{\rm s}^{-1}}\right)^2\quad
   ({\rm erg}/\lsun), \eqno(11)$$
where the empirical IMLR for the ICM has been used (with $h_{50}=1$), and
$Z^{\rm Fe}_{\rm w}$ is the average metallicity of the winds. In wind models
for elliptical galaxy formation (e.g. Arimoto \& Yoshii 1987) galactic winds
are established at the culmination of the metal enrichment process, and
their metallicity is about 
3 times the average stellar metallicity.
Therefore, by decreasing $<\! Z^{\rm Fe}_*\!>$
by a factor of 5 also $Z^{\rm Fe}_{\rm w}$ 
is seemingly decreased, and resulting
galactic wind heating of the ICM is increased by the same factor.

Galactic wind heating has great importance for the evolution of the ICM,
and therefore for the X-ray evolution of clusters of galaxies
(Kaiser 1991; Cavaliere, Colafrancesco, \& Menci 1993; Metzler \& Evrard 1994). 
Here we just emphasize that a reduction
of the estimated metallicity of stars in ellipticals automatically implies
an increase by the same factor in the estimated heating of the ICM.

\medskip
\centerline{\it 3.3.3. The Failure of the Standard Chemical Model?}
\medskip
Further challenge to the SCM comes  from {\it ASCA}
observations
of the ICM in several clusters of galaxies. Preliminary results
appear indeed to indicate an enhancement of the $\alpha$ elements (notably
neon) relative to iron in the ICM, which argues for a dominant Type II SN
enrichment (Mushotzky 1994; Mushotzky et al. 1996). 
When applied to a cluster as a whole, the SCM
predicts near solar $\alpha$/Fe ratios when the total mass of each element
is considered, i.e., the total amount locked in stars within galaxies plus the
amount diffused in the ICM (RCDP). The model also predicts an $\alpha$-element
enhancement compared to iron in the stellar component of galaxies, and
instead an $\alpha$-element depletion 
in the ICM. This {\it chemical asymmetry} -- the signature
of the SCM -- follows from the promptly released SN II products being
predominatly locked into stars during the intense star forming
activity at the formation epoch of ellipticals, while a main fraction of the
total iron -- more slowly released by SNIa's -- flows out of galaxies
into the ICM after galactic winds are established and most star formation
has ceased.
As already mentioned, optical data suggest that Mg is indeed enhanced in
the stellar component of ellipticals, which is in agreement with the
expectations from the SCM. However, the enhanced [$\alpha$/Fe] ratio
in the ICM of several clusters (up to $\sim 2$ times solar according
to Mushotzky et al. 1996) appears at variance with what predicted by
the SCM, although a reconciliation might not be completely impossible
(Loewenstein \& Mushotzky 1996; Ishimaru \& Arimoto 1996).

In summary, the abundances derived from X-ray data relative to the ISM of
individual ellipticals require SNIa's to be irrelevant at the present time,
and the X-ray-derived abundance ratios for the ICM require SNIa's to have been
irrelevant in the past. In other words, all X-ray data appear
consistent with the
notion that all elements in clusters of galaxies, including iron, are produced
by Type II SNs, with a negligible contribution from SNIa's (see also
Loewenstein \& Mushotzky 1996). Thus
the SCM dramatically fails to account for the chemical abundances as derived
from the X-ray observations.

The production of all the iron that is observed in clusters of galaxies may
indeed be produced by Type II SNs alone, suffice the IMF to be somewhat
flatter than in the solar neighborhood, and an IMF  slope $x\simeq 0.9$ would
be sufficient for this purpose (Ciotti et al. 1991; Arnaud et al. 1992;
RCDP; Matteucci \& Gibson 1995). A variant of this scenario appeals to bimodal star formation, with the
starburst activity in early ellipticals generating only stars more
massive than
$\sim 2$ or 3 $\msun$, as advocated by Arnaud et al. (1992) and Elbaz,
Arnaud, \& Vangioni-Flam (1995). 
\medskip
\centerline{\it 3.3.4. Suppressing Binary Star Formation in Ellipticals?}
\medskip
There are therefore viable alternatives to SNIa's for the producion of all the
iron that is observed. This is not the question. The question is instead why
SNIa's should
play a major role within the Galaxy, and yet be irrelevant in the ecology of
ellipticals and clusters of galaxies. This indeed requires SNIa's to be
selectively suppressed in the stellar populations of clusters of galaxies
(ellipticals), with respect to spirals.
This embarrassing implication is discussed next.

SNIa's are believed to be the product of binary star evolution, in which a
white dwarf accretes material from a companion until a thermonuclear runaway
is ignited in the white dwarf. Suppressing by a great factor SNIa's in
elliptical compared to spiral galaxies therefore translates
into the request of suppressing binary star formation by nearly the
same factor.
Binary star frequency in ellipticals can be estimated in two ways. As
already
mentioned the hard spectral component is generally ascribed to a population of
low mass X-ray binaries (LMXRB's), 
and there appears to be no shortage of such objects  in
elliptical galaxies judging from the ratio of the X-ray luminosity of
the hard component
to the optical luminosity being  the same as in spirals (Canizares et al.
1987; Pellegrini \& Fabbiano 1994; Pellegrini 1994;
Matsushita et al. 1994).

Another manifestation of binaries is represented by nova outbursts. In
Virgo ellipticals the
frequency of novae per unit galaxy light is  the same
as in nearby spirals (Della Valle et al. 1994). Again, this argues
for the binary frequency in ellipticals being the same as in spirals.
We conclude that the option of a drastically suppressed binary star
formation in ellipticals does not find independent support. Apparently,
suppression should apply only to the particular kind of binary systems that
are able of triggering a SNIa event. SNIa precursors belong certainly to
different kinds of binaries compared to LMXRBs or Novae. LMXRBs contain a
neutron star, i.e., had a massive star primary. Novae have a white
dwarf primary
and an unevolved low mass secondary. In SNIa precursors instead, also the
secondary component must have evolved away from the main sequence. Although
a difference exists after all, it is hard  to find arguments supporting
a selective suppression of just those binaries with a primary star less massive
than $\sim 8\,\msun$, and a secondary star more massive than $\sim 0.8\,\msun$,
while keeping unchanged the productivity of other binaries.
The first limit ensures that the primary will become a white dwarf, the second
one that the secondary will evolve off the main sequence in less than $\sim
15$ Gyr. We conclude that SNIa precursor-binary star suppression offers a
formal solution to the dilemma, but a very contrived one indeed.

It is worth emphasizing that bimodal star formation
does not solve the problem. If most of the metals in clusters of
galaxies are produced in early starbursts where only stars more massive than
$\sim 2\,\msun$ are formed (Arnaud et al. 1992), then this starburst population
will also include all the SNIa precursors as well. The total mass of the binary
SNIa precursors is in fact in the range from $\sim 4$ or 5 $\msun$ up to $\sim
16\,\msun$, with a mass of the primary in the range between $\sim 3$ to $\sim 8
\,\msun$, irrespective of the specific SNIa model (Greggio \& Renzini 1983a;
Iben \& Tutukov 1984). Indeed, there appears to be no way of
triggering a destructive thermonuclear explosion in binaries with
initial total mass below $\sim 4\msun$.
Bimodal star formation models for the chemical
enrichment of clusters of galaxies explicitly assume that only stars with
$M\gsim 2\,\msun$ are formed during early bursts, but in addition implicitly
assume the binary progenitors of SNIa's to be completely suppressed in the
burst population. Formally, this extra-assumption could be avoided if the burst
population were to produce only individual stars more massive than $\sim
8\,\msun$, either single or in a binary. If so, the population would not
produce carbon-oxygen white dwarfs at all, and the necessary objects to make
SNIa's would be absent in the population. Given the arbitrary nature of the
bimodal star formation hypothesis, and in particular of the value of the lower
mass cutoff, an assumption of this kind is certainly not less legitimate
than others; it would actually have the advantage of avoiding the ad hoc
assumption of the binary suppression in the burst population.

A simple chemical evolution model can be used if SNIa's are indeed
irrelevant in the iron budget of ellipticals and clusters.
In this case, in the closed-box approximation the chemical evolution
of an elliptical galaxy is well described by:
$$ <Z^{\rm Fe}_*> = y^{\rm Fe} - {M_{\rm gas} Z^{\rm Fe}_{\rm gas} \over M_*},
          \eqno(12)$$
where $y^{\rm Fe}$ is the {\it yield} of iron produced by each stellar
generation, and 
$M_{\rm gas}$ and $Z^{\rm Fe}_{\rm gas}$ represent mass and iron
content of the ISM at any given time, repsectively (e.g. Tinsley
1980). Therefore, when a
wind is established the amount of iron that is ejected ($M_{\rm gas}
Z^{\rm Fe}_{\rm gas}$) can be obtained by this relation, and the resulting
IMLR of the ICM turns out to be :
$$ \left({M^{\rm Fe}_{\rm SNII}\over \lb}\right)^{\rm ICM} = (y^{\rm Fe} -
         <Z^{\rm Fe}_*> ) {M_* \over \lb}. \eqno(13)$$
In this purely chemical
framework, a low iron abundance in the stellar component helps getting
a higher IMLR in the ICM, as a larger fraction of the iron yield
remains available for ICM enrichment after star formation has ceased.
Following Nomoto et al. 
(1993), we adopt the iron yield $y^{\rm Fe}$ by massive stars as
$1.4 Z^{\rm Fe}_\odot$ for $x=1.05$
and $0.21 Z^{\rm Fe}_\odot$ for $x=1.50$. Again we recover the result
that a fairly flat IMF ($x\lsim 1$) is required to account for the iron
in the ICM,
while no iron at all would be left for polluting the ICM in the case
$x\gsim 1.5$, even if the iron abundance in stars is as low as 0.2 solar.
We recall that a flat IMF would also help producing synthetic stellar
populations  as red as the observed ellipticals (Arimoto \& Yoshii 1987).

\medskip
\centerline{\bf 4. WAYS OF HIDING IRON TO X-RAY OBSERVATIONS}
\medskip

So far we have assumed that the iron ejected by mass losing stars and SNs
is neither further diluted nor astrated from the ISM. 
This is equivalent to
assume that the abundance indicated by equation (1) is indeed the one to be
compared to the ISM abundance as measured from X-ray observations. This
assumption is now relaxed in the attempt of exploring if a solution to the
iron discrepancy can be found in this direction.

\medskip
\centerline{\it 4.1 Is Iron Mostly in Flakes?}
\medskip

When comparing the predicted abundance from equation (1) to abundances derived
from X-ray observations one implicitly assumes that all the iron ejected
by stars and SNIa's is now in the gas phase of the ISM. Actually, it is likely
that some fraction of the iron condenses  into dust grains
(or {\it iron flakes}) while in the circumstellar envelope of mass losing red
giants, and remains locked in them for some time after ejection. 
Indeed, along with other refractory elements, iron appears to be
depleted in planetary nebulae by factors up to $\sim 100$ (Clegg 1989).
Seemingly, part of the fresh iron synthetized by SNIa's could also
condense into solid particles when heating from the Ni and Co
radioactive decay
has sufficiently declined.
Such solid iron could join the hot ISM phase
only after sputtering has dissolved  the grains.

While there is little doubt that at least a fraction of the iron is
injected in the
form of solid particles, the question is as to whether iron remains in a solid
phase for a time long enough  to account for the iron discrepancy.
The time required to evaporate 90\% of iron particles in a $10^7$ K plasma
is $\sim 10^5/n_{\rm e}$ yr cm$^{-3}$ (Itoh 1989). Electron densities
in the ISM of ellipticals range from $\sim 10^{-1}$ near to center to $\sim
10^{-3}$ several effective radii away, and therefore evaporation times range
from $10^6$ to $10^8$ yr. This compares to a flow time of a few Gyr, and we
conclude that only a tiny fraction of iron could be hidden in dust particles,
thereby subtracting itself to X-ray observations.

\medskip
\centerline{\it 4.2. Or in Super Iron-Rich Jupiters?}
\medskip

Here we mention yet another, admittedly exotic solution to the iron discrepancy.
Thermal instabilities causing a fraction of the gas
to drop out of the flow locally have often been invoked as the ultimate
depository for both cluster and galaxy inflows (e.g., Fabian et al. 1986;
Sarazin 1986). Together with this hypothesis comes
the additional conjecture that such instabilities would lead to the formation
of low mass,  unobservable objects
such as lower main sequence stars, brown dwarfs, or ``Jupiters''.
This scenario is far from having been supported by independent observations,
while it has encountered several physical difficulties (e.g., 
Murray \& Balbus 1992; Kritsuk 1992). Here we just adopt
it as a working hypothesis.

The sites of SNIa explosions remain as local enhancement of the iron abundance
before Rayleigh-Taylor instabilities and mixing homogenize the ISM. It has
been suggested that such inhomogeneities may work as seeds for the growth
of local thermal instabilities, thanks to their high iron enhancement
causing more efficient cooling (Mathews 1990). Therefore, one may speculate
that most of the iron released by SNIa's could ultimately be disposed in the
dark residual of the thermal instabilities, e.g., in super iron-rich Jupiters,
rather than enrich the ISM and the ICM.

By good fortune, this scenario has one testable prediction. Local thermal
instabilities -- if they exist -- should be more efficient in dropping SNIa
products out of the flow near the center where the cooling time is short,
compared to the outer regions. Thus, iron should be preferentially depleted
near the center, and a positive abundance gradient should develop.
Actual observations seem in case to indicate the opposite gradient, with iron
being more enhanced near the center (Mushotzky et al. 1994). We conclude that
hiding iron into jupiters does not appear to be a viable solution to the
discrepancy.

\medskip
\centerline{\it 4.3. Is the ISM Diluted by the ICM?}
\medskip

Equation (1) assumes that the whole ISM in a galaxy is made of materials that
have been lost by the stellar component of the galaxy itself. 
Elliptical galaxies are often inbedded in a hot ICM, and  ICM
materials can be accreted if an inflow is
established, thus diluting the indigenous ISM
(RCDP). For the specific case of NGC 1399 -- the cD galaxy in the Fornax
cluster -- RCDP's detailed flow modelling indicates that a dilution by
up to a
factor of 3 is possible, thus easing the discrepancy somewhat. RCDP also argue
that some dilution may also be effective in the case of NGC 4472, that is part
of a subcondensation within the Virgo cluster.

The hypothesis that accretion from the ICM and dilution are responsible for the
iron discrepancy has some testable consequences. The first is that accreting
galaxies should be overluminous in X-rays, as the additional $PdV$ work on the
infalling material needs to be radiated away in a quasi-stationary cooling flow.
NGC 1399 is in fact somewhat more luminous than expected from an isolated
inflow regime (RCDP), which apparently supports the  dilution hypothesis.
However, one
expects iron dilution  to be more effective  in galaxies that are
located in a cluster or group with a dense ICM, 
and less effective 
in isolated ellipticals, that are not
projected on a high X-ray surface brightness environment. 
Among the galaxies in Table 2, some are member of the Virgo cluster 
-- and dilution may be invoked -- 
but others (e.g., NGC 4636 and NGC 720) are isolated and in a low
density environment. Yet, the iron abundance in the latter galaxies is as low
-- or even lower -- than that of cluster members. We conclude that
dilution does not offer a viable solution either.

\medskip
\centerline{\bf 5. IRON-L DIAGNOSTICS}
\medskip

\medskip
\centerline{\it 5.1. XSPEC vs Meka vs Masai}
\medskip

The abundances reported in  \S 2.2 have been obtained from the 
standard RS  thin plasma emission model incorporated in the XSPEC package, and
in this section we compare these results with those obtained using
the so-called Meka model (Mewe, Gronenshild \& van den Oord 1985; 
Mewe, Lemen, \& van den Oord 1986; Kaastra 1992) and Masai model
(Masai 1984).
Such models use somewhat different  input physics, e.g., 
ionisation, collisional excitation and recombination rates, energy
levels and equivalent widths, etc.
At $T\simeq $1 keV, iron ions have still several bound electrons, 
and therefore spectra of the iron L-line blend are very complex.
Features of the iron-L blend can  differ considerably in different
models, and so do the resulting iron abundances.

Apart from the use of different models, we then proceeded
in exactly the same way as descibed in \S 2.2.
For the solar iron abundance we have used  $4.68\times10^{-5}$ by number for
all models, and the results are shown in Table 3 and Fig. 4.
The physical parameters obtained by these models are quite different. 
Temperatures derived from Meka model are  about 0.2 keV and 0.15 keV
lower than those derived from the RS model  and the Masai model,
respectively. The iron abundance can differ sometimes by a factor 2-3
from model to model. 
The Meka model gives  iron abundances similar to  
that of the RS model, except for NGC1399 to which the Meka model 
gives slightly lower iron abundance. We note that the RS model
gives the highest temperature for this galaxy.  
The Masai model gives systematically larger iron 
abundance than the RS  model.
For the three X-ray luminous galaxies in Virgo
(NGC4406, NGC4472 and NGC4636) the abundances are almost solar.
Even the Masai model, however, gives abundances  $\sim 0.2-0.3$ solar
for NGC720, NGC1404, and NGC4374, with an upper limit $\sim$0.6 solar.

In conclusion, the size of the iron discrepancy is somewhat reduced,
but by no means eliminated, when adopting iron abundances from the
Masai model. This is especially so when allowance is made for the 
iron enrichment by SNIa's.

\medskip
\centerline{\it 5.2. The Iron-L Problem}

The iron-L blends contain a few hundreds of lines produced by several 
ionization stages, and the specific blends depend on temperature.
The atomic physics of such multi-electron ions is rather complicated,
hence affected by uncertainties that are difficult to quantify, but
that may be rather large. Atomic data such as ionization, recombination,
and especially collisional excitation rate coefficients are still 
controversial, which accounts for the differences among the various models.
New calculation of plasma spectra
using new atomic data (Liedahl, Ostenheld, \& Goldstein 1995)
show that the total power of the iron-L shell lines
can be up to 40\% different from Meka model even using
the same ionization and recombination rates. The temperatures
explored  in this study ($kT\ge 1.2$ keV) are more pertinent to galaxy
groups and clusters, where problems with the iron-L diagnostics have
also been encountered (Fabian et al. 1994), and therefore one cannot firmly
exlude even larger discrepancies at the lower temperatures that are typical
of elliptical galaxies.

As shown in the previous section, the derived abundances from different
models may differ by a  factor 2 or 3.
When fitting spectra, besides abundance the fit parameters  include
two plasma temperatures, the strengths of these two components, and
the hydrogen column density of the cold gas
absorbing soft X-rays. The procedure is therefore much more complicated than 
a simple isothermal plasma model, and so is the circulation of the errors
affecting the derived abundance.

To assess  the dependence  of the resulting abundance on the model
ingredients
we have calculated spectra of an isothermal plasma with 
{\it half} solar abundance
using the RS model, convolved them with the {\it ASCA} response, and
then fitted the resulting spectra
with isothermal Meka and Masai models. 
Photoelectric absorption was not included in this exercise.
When $kT\gsim$1 keV  or $\lsim$0.5 keV all models agree fairly well,
but for $0.5\lsim kT\lsim 1$ the abundances
from Meka and Masai models are larger  by a
factor 1.4 and 1.8 than those from the RS  model, respectively.
For example, for a $kT=0.9$ keV RS spectrum the abundances from the
Meka and Masai models are 0.7  and 0.9 solar,
respectively. 

It is worth tracing the origin of these differences. The excitation
rate of iron levels is the same in Masai and  Meka models, while
differences exist in oscillator strengths, branching ratios,
and especially ionization and recombination rates. 
When using the same ionization and recombination rates the
calculated spectra of these two models are almost identical, with 
differences of at most $\sim$30\% for some strong lines.
However, when  spectra simulated with the  Meka model are fitted with
the Masai model,
the latter gives abundances which are a factor $\sim$1.5 larger than
those in the Meka model. Much larger differences are encountered in
more
complex situations, e.g., for two temperature plasmas including
photoelectric absorption.

The accuracy of the iron abundances obtained from the iron-K complex
at $kT\simeq 6.7$ keV is generally much higher than that of abundances from
the iron-L complex, as the atomic physics is much simpler for the 
hydrogen-like and helium-like configurations producing the iron-K blend.
So, when both iron-K and iron-L features are sufficiently strong in a
given object, abundances from the iron-K lines can be used to asses 
the accuracy of the iron-L diagnostics. Unfortunately, for $kT\simeq 1$ keV 
the iron-K lines are very weak and cannot be detected, but for
$kT\simeq 2-3$ keV both the iron-L and iron-K lines are detectable. 
This is the case for the
ICM of the Virgo cluster, and in this mood we have analyzed the {\it
ASCA} X-ray spectrum (from the GIS detector) of a nearly isothermal
region about 10 arcmin from the center of M87.
The isothermal RS model indicates $kT\simeq 3$ keV and  an iron
abundance $\sim$0.3 solar for this region. When the spectral range
around the iron-K complex is excluded from the fit the RS model gives
$kT=2.89\pm 0.07$ keV and an iron abundance (now determined only from
the iron-L lines)  $0.41(0.34-0.50)$ solar. When excluding the
spectral range below 1.8 keV the model gives 
$kT=2.90 \pm 0.1$ keV and an
iron abundance (now determined only from the iron-K complex)  
$0.40(0.34-0.47)$ solar.
Thus, for $kT\simeq 3$ keV the iron-K and iron-L iron abundances 
agree fairly well. However, this does not ensure the adequacy of the
current iron-L diagnostics at lower temperatures, as for $kT\lsim 1$ keV 
the iron-L lines are produced by iron ions in lower
ionization stages, hence by completely different transitions, 
and the atomic physics involved is correspondingly much more
complicated, hence more uncertain. Meanwhile, at such low temperatures
the iron-K complex disappears and no comparison between iron-K and
iron-L diagnostics is possible.

Besides possible systematic errors due to the atomic
physics ingredients, the derived iron abundances may also be compromized by the
assumption of an isothermal ISM.
SNIa ejecta could be hidden as an un-mixed super-hot phase in the ICM.
When this assumption is relaxed, and a multi temperature ISM is
allowed the derived iron abundances may differ considerably (Fabbiano, 1995).
To explore this  effect the spectra of the seven galaxies in Table 2
have been  fitted with a two-temperature RS model plus an  hard
component representing the LMXRB contribution, assuming the  abundance
of each ISM phase to be  the same. The resulting temperatures of the
two components cluster around $kT\simeq 0.3$ and 1 keV, 
and the derived iron abundances increase by at most $\sim 30-40$ \%,
while the reduced $\chi^2$
does not decrease much, exept for NGC4406 and NGC4636.
All in all, the derived iron abundances remain below solar.
We conclude that ISM temperature inhomogeneities alone are unlikely to
solve the iron discrepancy.

\medskip
\centerline{\it 5.3. Iron-L in Other Astrophysical Objects}
\medskip
Further atomic physics studies may help assessing whether sizable
errors were affecting the atomic parameters used in the various thin
plasma emission models. However, a more direct, empirical assessment
may come from iron-L abundance determinations of astronomical objects
whose iron abundance is independently known. In this section, we present
a preliminary attempt in this direction. It is actually difficult to
find individual objects with both well determined optically and X-ray
determined iron abundances. For most objects in this section one may 
argue the circumstantial evidence may favor near solar abundance.
For others, e.g., galaxy groups and lose clusters, one may argue their
iron content should not greatly differ from that of rich clusters.

\centerline{\it 5.3.1. Binary and Stellar X-Ray Sources}
\medskip

Strong emission lines are observed during the eclipse of close binary
systems. In particular, the eclipse spectrum of Vela X-1 exhibits
intense K lines from Fe, Mg, Si, S, and Ar (Nagase et al. 1994). 
However, as mentioned above, the low fluorescent yield for iron-L
lines hampers the significant detection of these lines from compact
X-ray sources. {\it ASCA} measurements of stellar X-ray sources
$\beta$ Cet and $\pi^1$ UMa (Drake et al. 1994) give the iron
abundance $0.77^{+0.24}_{-0.18}$ and $0.41^{+0.24}_{-0.10}$ solar,
respectively, while the photospheric iron abundances for these stars
obtained by optical measurements are $0.80-0.89$ solar (Lambert \& Ries 1981;
Kovacs, 1982) and 0.54 solar (Hearnshaw 1974), respectively.  
RS CVn system (AR Lac) suggests that the metallicities 
including iron-L features are a factor of 3-4 below the solar values 
(White 1996). For AR Lac, 
Singh et al. (1996) have derived the iron abundance
0.53 solar with Meka model by excluding the iron-L ($0.7-1.74$ keV) region.
This value is consistent with the optical measurement by Naftilan \& 
Drake (1977) who gave 0.5 solar, 
but once the iron-L region is included, the resulting iron
abundance becomes as low as 0.29 solar (Singh et al. 1996).

\centerline{\it 5.3.2. Supernova Remnants}
\medskip

Supernova Remnants are the rich source of emission lines. In fact {\it ASCA}
has detected remarkable spectra which are dominated by strong emission
lines, e.g., from Cas-A (Holt et al. 1994)
and W49B (Fujimoto et al. 1995). In connection to the iron-L
problem in elliptical galaxies the interesting objects are the SNRs
with $kT \simeq
1$ keV. Recent results for three young SNRs 
(0509-67.5, 0519-69.0, N103B) in the LMC with
$kT=1.1-1.5$ keV are reported by Hughes et al. (1995). The
observed spectra exhibits strong iron-L lines and K lines from Si, S,
Ar and Ca. Lines from O, Ne, Mg are relatively weak. They show that,
qualitatively, the spectra agree well with the W7 model
by Nomoto et al. (1984) for SNIa. Again, the rough
consistency with the model calculation including the iron-L lines with
the observed data suggests that the model estimation should not differ
by more than a factor of 2, at least for $kT\gsim 1$ keV. However, the
iron abundance in the Cygnus loop, an older and cooler SNR, appear to be
appreciably subsolar ($\lsim 0.1$ solar, Miyata et al. 1994),
although its low temperature $\sim 0.3$ keV suggests that the objects 
could be out of ionization equilibrium.

\centerline {\it 5.3.3. Starburst Galaxies and AGNs}

{\it ASCA} observations of starburst galaxies indicate that the X-ray
emission from these systems is not explained by a single component. 
There seems to be at least three emission components, as best demonstrated
for the brightest galaxy M82 (Tsuru et al 1994). The observed energy
spectrum shows very weak iron-K lines, He-like and H-like Si and Mg lines,
and the iron-L complex. The ratio of H to He-like Si lines indicates
a plasma temperature of $1 \sim 1.3$ keV, which is too high to
account for the ratio of H- to He-like Mg lines, 
which rather suggests a plasma temperature less than $\sim$1 keV. 
This cool component may extend out
of the galaxy, judging from the X-ray morphology taken by {\it ASCA}.

This complex situation makes starburst galaxies rather unsuitable to
calibrate the iron-L diagnostics. However, a preliminary value of 
$\sim 0.1$ solar is suggested for M82 (Tsuru et al. 1994) 
and NGC 253 (Awaki et al. 1996).

AGNs (in particular Seyfert galaxies) commonly exhibit fluorescent
iron-K lines. Highly obscured objects emit strong K lines from
Fe, Mg, Si, Ar and S as a result of fluorescence in the
surrounding medium irradiated by a strong continuum X-ray source that
is centrally located.
However, since the fluorescence yield for
iron-L lines is only about 0.4\%, these lines have not been significantly
observed from the scattering medium. Iron-L lines are seen in the
spectrum of Seyfert-2 galaxies in which AGN and starburst activities
occur simultaneously (for example NGC 1068, Ueno et al. 1994). 
The iron-L feature suggests
a thermal plasma with $kT \sim 0.6$ keV and the resulting iron abundance is
about 0.3 solar. Therefore the situation is
similar to that of starburst galaxies.

\centerline{\it 5.3.4. Groups and Clusters of Galaxies}
\medskip

The iron-L lines have been observed in clusters of galaxies in two broad
circumstances: in poor clusters and groups with relatively cool ICM
($kT\lsim 3$ keV), and in the central cool regions of large, 
{\it cooling flow} clusters. The former case includes NGC5044 group,
HCG51 (Fukazawa et al. 1996), and Fornax cluster (Ikebe 1996). The
ICM temperatures are typically $kT\gsim 1$ keV, and the iron abundance
determined from the iron-L lines is 0.3 -- 0.4 solar. These abundance
values agree with those determined for rich clusters (with $kT\gsim 3$
keV) from the iron-K lines, which suggests that the iron abundance may 
be the same in different classes of clusters (Ohashi 1995). This
agreement seems to indicate that the iron-L diagnostics is correct, at
least at the temperature of these clusters.
 
However, the situation is different for cooler clusters and groups. 
For example,
an iron abundance as low as 0.06 solar was derived for the NGC 2300
group (Mulchaey et al. 1993; Sakima, Tawara, \& Yamashita, 1994) 
whose ICM temperature is $0.9$ keV, and a
seemingly low abundance has been estimated for the group HCG62 
(cf. Ohashi 1995). The resulting IMLR is exceptionally low for NGC
2300, which has forced RCDP to appeal to a
rather contrived scenario, with the group having first suffered extensive gas
and iron losses, then followed by reaccretion of matter extensively
diluted with pristine material. Fig. 5 (adapted from Renzini 1996c),
shows the iron abundance of the ICM clusters and groups as a function
of
their temperature. It is apparent that for $kT\gsim 3$ keV the iron
abundance (derived for the iron-K feature) is fairly constant, a well
known result. However, for lower temparatures iron-L features are
used, and the ICM iron abundance starts showing large variations that
appear
systematic with temperature, rather than random. For $T$ decreasing
from $\sim3$ keV to $\sim 1$ keV the abundance appears to increase,
from $\sim 0.3$ solar to $\sim$ solar. Below $\sim 1$ keV, however,
the abundance drops percipitously to virtually zero (the lowest points
in Fig. 5 are actually upper limits at 0.01 solar!), with a tight
abundance-temperature correlation. This systematic
trend with temperature may find and astrophysical explanation, but it
would be indeed more naturally understood as the result of a
systematic bias of the iron-L diagnostics, with iron being
overestimated for $kT\sim 1$ keV (when Fe XXIV-XXV dominate), and 
progressively more and more
underestimated at lower temperatures, when lower and lower ionization
stages (down to Fe XVI) become responsible for the iron-L emission. 
Actually, this
trend may provide some hint as to which aspect of the atomic physics
may be called into question.

An interesting case of a rich cluster
with a cool core is the Centaurus cluster
(Fukazawa et al. 1994, Ikebe 1996). The temperature of the widely
spread ICM is 3.8 keV, but the energy spectrum in the
central $\sim300$ kpc region shows clear presence of cooler, $\sim 1$ keV gas
characterized by strong iron-L lines. The spectral fit indicates that
the metal abundance of both hot and cool gas is consistent with being the
same, both rising to about 1.7 solar in the center from the 0.3 solar
in the cluster outskirts. However, the 2-component (or multi-component) nature
of the spectrum makes the independent abundance determination using
only iron-L lines difficult. In fact, if one lets $kT$ and abundance
of the cool component to vary freely, they remain  undertermined.
Nonetheless, the consistency of the abundance with the hot component 
suggests that the iron-L lines diagnostics may not be affected
by  large systematic errors, at least at the temperature such cool
component ($kT$ slightly above 1 keV).

\centerline{\bf 6. DISCUSSION AND CONCLUSIONS}
\medskip
The iron abundance  in the ISM of elliptical galaxies -- 
as derived from X-ray observations of the iron-L complex -- appears to
be as low as 0.1--0.4 solar. Such low abundances are at
variance with the abundance expected from the stellar population as
derived from current population synthesis methods
of the optical spectrum, especially if allowance is made for the
further enrichment by SNIa's at empirically determined rates, which
together
imply an abundance $\gsim$ 2 solar.
This strong discrepancy appears to shake our understanding of supernova
enrichment and chemical evolution of galaxies.

If the ISM iron abundances indicated by X-ray observations are
correct, then not only population synthesis methods should be in
error, but either the empirical SNIa rate in ellipticals or the
current theoretical 
estimates of the iron yield of these SNs should also be in error.
Moreover, the iron share between the ICM and cluster galaxies would
change from being nearly equal, to most of iron being in the ICM, with
the
ICM containing $\sim 5/6$ of the total iron and stars in galaxies only
$\sim 1/6$. This would imply a dramatic mass loss from galaxies at
early times, and along with it a very strong heating of the ICM that
would have major effects on the evolution of the ICM itself.

Formally, the X-ray observations seem to require the SNIa activity to
be  virtually
suppressed  in cluster ellipticals, compared to spiral galaxies.
However, we point out that binary stars indicators such as the Nova
rate and the hard X-ray to optical luminosity ratio in ellipticals
suggest the binary star population to be virtually the same as in
spiral galaxies. Other astrophysical explanation of the discrepancy,
such as hiding iron in solid particles or in Jupiter-like objects, or
diluting it with an iron-poor ICM do not appear to be promising.

Having failed to find an attractive astrophysical solution, we have 
explored the opposite option, scrutinizing in some detail the tool
used in deriving iron abundances from X-ray spectra. It is found that
three different thin plasma models give basically the same iron
abundances when applied to astrophysical thin plasmas with $kT\gsim 3$
keV. However, for lower temperatures, and especially for $kT\lsim 1$
keV, different codes can give iron abundances that differ by as much
as a factor of three. The complex nature of the iron ions responsible
for the iron-L emission around 1 keV is probably to blame for these
discrepancy, as the atomic physics parameters of these ions are poorly
known. This suspicion is somewhat reinforced by several empirical
results, indicating iron abundances systematically below solar  when
temperatures lower than $\sim 1$ keV are encountered. This is the case
of objects as disparate as binary X-ray stars such as RS CVn's and
Algols, SN remnants, starburst galaxies, and galaxy groups. The latter 
is an especially clear case: the intragroup iron abundance seems to drop by
two orders of magnitude for $kT\lsim 1$ keV. 

All these evidences,
coupled to the failure of our attempts to find a reasonable
astrophysical explanation, suggest to us worth exploring further
the possibility that the very low iron
content derived from X-ray observations  may
be an artifact of the thin plasma model tools employed in the
diagnostics of the iron-L X-ray emission complex. While we believe
worth continuing to critically reconsider our views of the enrichment
process of galaxies, groups and clusters, we also believe worth
extending such a critical attitude to X-ray diagnostic tools.

\centerline {\bf Acknowledgement:} We are grateful to our referee,
Michael Loewenstein, for his useful and detailed constructive 
suggestions. This work was financially supported
in part by Grant-in-Aids for the Scientific Research (NA; No.06640349 and 
07222206) by the Japanese Ministry of Education, Culture, Sport, and
Science.
AR gratefully acknowledges the support of the Japan Society for the
Advancement of Science (JSPS), that made possible an extended visit to
the University of Tokyo during which this project was started.
KM and YI acknowledge financial supports from the JSPS.

\vfill\eject

\centerline{\bf REFERENCES}
\bigskip\pn
\ref{Abia, C., Canal, R., \& Isern, J. 1991, ApJ, 366, 198}
\ref{Arimoto, N., \& Yoshii, Y. 1987, A\&A, 173, 23}
\ref{Arnette, W.D., Branch, D., \& Wheeler, J.C. 1985, ApJ 295, 589}
\ref{Arnaud, M., Rothenflug, R., Boulade,O., Vigroux, R., \&
    Vangioni-Flam, E. 1992, A\&A, 254, 49} 
\ref{Awaki, H. et al. 1996, in UV and X-ray Spectroscopy of Astrophysical
    and Laboratory Plasmas, ed. Y. Yamashita and T. Watanabe,(Universal
Academy Press) ,p327}
\ref{Awaki, H. et al. 1994, PASJ, 46, L65}
\ref{Awaki, H., Koyama, K., Kunieda, H., Takano, S., Tawara, Y., \& Ohashi, T.
     1991, ApJ, 366, 88}
\ref{Barbuy, B. 1992, in The Stellas Populations of Galaxies, ed. B. Barbuy \&
     A. Renzini \klu 143}
\ref{Baum, W.A., Thomsen, B., \& Morgan, B.L. 1986, ApJ, 301, 83}
\ref{Bessell, M.S., Sutherland, R.S., \& Ruan, K. 1991, ApJ, 383, L71}
\ref{Bica, E. 1988, A\&A, 195, 75}
\ref{Binney, J., \& Tabor, G. 1995, MNRAS, 276, 663}
\ref{Boroson, T.A., \& Thompson, I.B. 1991, AJ, 101, 111}
\ref{Branch, D., Livio, M., Yungelson, L.R., Boffi, F.R., \& Baron, E.
     1995, PASP, 107, 1019}
\ref{Buzzoni, A., Gariboldi, G., Mantegazza, L. 1992, AJ, 103, 1814}
\ref{Canizares, C.R., Fabbiano, G., \& Trinchieri, G. 1987, ApJ, 312, 503}
\ref{Cappellaro, E., Turatto, M., Benetti, S., Tsvetkov, D. Yu., 
     Bartunov, O. S., \& Makarova, I. N. 1993, A\&A, 268, 472}
\ref{Carlberg, R. 1985, ApJ, 286, 404}
\ref{Carollo, C.M., \& Danziger, I.J. 1994a, MNRAS, 270, 523}
\ref{--------------. 1994b, MNRAS, 270, 743}
\ref{Cavaliere, A., Colafrancesco, S., \& Menci, N. 1993, ApJ, 415, 50}
\ref{Ciotti, L., \& Ostriker, J. 1996, in preparation}
\ref{Ciotti, L., D'Ercole, A., Pellegrini, S., \& Renzini, A. 1991, ApJ, 376,
     380}
\ref{Clegg, R.E.S. 1989, in Planetary Nebulae, ed. S. Torres - Peimbert
     (Dordrecht: Kluwer), p. 139}
\ref{Couture, J., \& Hardy, E. 1988, AJ, 96, 867}
\ref{David, L.P., Forman, W., \& Jones, C. 1990, ApJ, 359, 29}
\ref{David, L.P., Jones, C., Forman, W., \& Daines, S. 1994, ApJ, 428, 544 }
\ref{Davidge, T.J. 1992, AJ 103, 1512}
\ref{Davies, R.L., Sadler, E.M., Peletier, R.F., 1993, MNRAS, 262, 650}
\ref{Davies, R.L. et al. 1987, ApJS 64, 581}
\ref{Delisle,S., \& Hardy, E. 1992, AJ, 103, 711}
\ref{Della Valle, M., Rosino, L., Bianchini, A., \& Livio, M. 1994, A\&A, 287,
     403}
\ref{de Vaucouleurs, G. 1948, Ann. d'Astrophys., 11, 247}
\ref{Donnelly, R.H., Faber, S.M., \& O'Connell, R.M. 1990, ApJ, 354, 52}
\ref{Dow, K.L. \& White, S.D.M. 1995, ApJ, 439, 113}
\ref{Drake, S.A., Singh, K.P., White, N.E., \& Simon, T. 1994, ApJ, 436, L87} 
\ref{Efstathiou, G., \& Gorgas, J. 1985, MNRAS, 215, 37p}
\ref{Elbaz, D., Arnaud, M., \& Vangioni-Flam, E. 1995, A\&A, 303, 345}
\ref{Fabbiano, G. 1995, in Fresh Views of Elliptical Galaxies, ed. A.
     Buzzoni, A. Renzini, \& A. Serrano, PASP Conf. Ser. 86, 103}
\ref{Fabbiano, G., Kim, D.-W., \&,Trinchieri, G. 1992, ApJS, 80, 531}
\ref{--------------. 1994, ApJ, 429, 94}
\ref{Faber, S.M. 1977, in The Evolution of Galaxies and Stellar
     Populations, ed. B.M. Tinsley, R.B. Larson, (New Haven: Yale
     University Observatory), p. 157}
\ref{Fabian, A.C., Arnaud, K.A., Bautz, M.W., \& Tawara, Y. 1994, ApJ,
     436, L63}
\ref{Fabian, A.C., Thomas, P.A., Fall, S.M., and White, R.E.,III 1986, MNRAS,
     221, 1049}
\ref{Forman, W., Jones, C., David, L., Franx, M., Makishima, K., \& Ohashi, T.
     1993, ApJ, 418, L55}
\ref{Franx, M., Illingworth, G., Heckman, T. 1989, AJ, 98, 538}
\ref{Freedman, W. 1989, AJ, 98, 1285}
\ref{Fujimoto, R., et al. 1995, PASJ, 47, L31}
\ref{Fukazawa, Y. et al. 1994, PASJ, 46, L141}
\ref{Fukazawa, Y. et al. 1996, PASJ, in press}
\ref{Gratton, R., \& Ortolani, S. 1986, A\&A, 169, 201}
\ref{Greggio, L. 1996a,in The Interplay between Massive Star Formation, 
  the ISM and Galaxy Evolution, 
  ed. D. Kunth, B. Guiderdoni, M. Heydari-Malayeri, T.X. Thuan, \&
  J.T. Thanh Van (Gyf-sur-Yvette: Edition Fronti\`eres), in press}
\ref{Greggio, L. 1996b, preprint}
\ref{Greggio, L., \& Renzini, A. 1983a, A\&A, 118, 217}
\ref{--------------. 1983b, Mem. SAIt, 54, 311}
\ref{Gorgas, J., Efstathiou, G., \& Arag\'on-Salamanca, A. 1990, MNRAS, 245, 
     217}
\ref{Hearnshaw, J.B. 1974, A\&A, 34, 263}
\ref{Holt, S.S., Gotthelt, E.W., Tsunemi, H., \& Negoro, H., 1994,
     PASJ, 46, L151}
\ref{Hughes, J.P. et al. 1995, ApJ, 444, L81}
\ref{Iben, I.Jr, \& Tutukov, A.V. 1984, ApJS, 54, 335}
\ref{Ikebe, Y., 1996, Ph.D. thesis, University of Tokyo (RIKEN, IPCR CR-87)}
\ref{Ikebe, Y., et al. 1992, ApJ, 384, L5}
\ref{Ishimaru, Y., \& Arimoto, N. 1996, submitted to PASJ}
\ref{Itoh, H. 1989, PASJ, 41, 853}
\ref{Kaastra, J.S. 1992, An X-Ray Spectral Code for Optically Thin Plasmas
(Internal SRON-Leiden Report, updated version 2.0)}
\ref{Kaiser, N. 1991, ApJ, 383, 104}
\ref{Kim, D.-W., Fabbiano, G., \& Trinchieri, G. 1992,  ApJ, 393, 134}
\ref{Kodaira, K., Okamura, S., Ichikawa, S., Hamabe, M., \& Watanaba, M. 1990,
     Photometric Catalog of Northern Bright Galaxies, Institute of Astronomy, 
     University of Tokyo}
\ref{Kovacs, N. 1982, A\&A, 120, 21}
\ref{Kritsuk, A.G. 1992, A\&A, 261, 78}
\ref{Lambert, D.L., \& Ries, L.M. 1981, ApJ, 248, 228}
\ref{Larson, R.B. 1974, MNRAS, 169, 229}
\ref{--------------. 1976, MNRAS, 176, 31}
\ref{Liedahl, D.A., Osterheld, A.L., \& Goldstein, W.H., 1995, ApJ, 438,
     L115}
\ref{Loewenstein, M. et al. 1994, ApJ, 436, L75}
\ref{Loewenstein, M., \& Mathews, W.G. 1991, ApJ, 373, 445}
\ref{Loewenstein, M., \& Mushotzky, R.F. 1996, ApJ, in press}
\ref{Makishima, K., et al. 1996, PASJ, 48, 171}
\ref{Masai, K., 1984, Ap.Sp.Sci, 98, 367 }
\ref{Mathews, W.G. 1990, ApJ, 354, 468}
\ref{Matsushita, K. et al. 1994, ApJ, 436, L41}
\ref{Matteucci, F., \& Gibson, B.K. 1995, preprint}
\ref{Matteucci, F., \& Fran\c cois, P. 1992, A\&A, 262, L1}
\ref{Matteucci, F., \& Greggio, L. 1986, A\&A, 154, 279}
\ref{Metzler, C.A., \& Evrard, A.E. 1994, ApJ, 437, 564}
\ref{Mewe, R., Lemen, J.R., \& van den Oord, G.H.J. 1986, A\&AS, 65,
     511}
\ref{Mewe, R., Gronenschild, E.H.B.M., \& van den Oord, G.H.J. 1985, A\&AS,
     62, 197}
\ref{Mould, J.R. 1978, ApJ, 220, 434}
\ref{Miyata, E., Tsunemi, H., Pisarski, R., \& Kissel, S.E. 1994,
     PASJ, 46, L101}
\ref{Murray, S.D., \& Balbus, S.A. 1992, ApJ, 395, 99}
\ref{Mulchaey, J.S., Davis, D.S., Mushotsky, R.F., \& Burstein, D. 1993, ApJ,
     404, L9}
\ref{--------------. 1966, ApJ, 456, 80}
\ref{Mushotzky, R. 1994, in Clusters of Galaxies, ed. F. Durret (Gyf sur
     Yvette: Editions Fronti\`eres), p. 167}
\ref{Mushotzky, R.F., Loewenstein, M., Awaki, H., Makishima, K., Matsushita,
     K., \& Matsumoto, H., 1994, ApJ, 436, L79}
\ref{Mushotzky, R.F. et al. 1996, ApJ, in press}
\ref{Naftilan, S.A., \& Drake, S.A. 1977, ApJ, 216, 508}
\ref{Nagase, F. et al. 1994, ApJ, 436, L1}
\ref{Nomoto, K., Thielemann, F.--K., \& Yokoi, K. 1984, ApJ, 286, 644}
\ref{Nomoto, K., et al. 1993, in Elements and the Cosmos, ed.
     R.J.     Terlevich, B.E.J. Pagel, R. Carswell, \& M. Edmunds (Cambridge:
     Cambridge Univ. Press), p. 55}
\ref{Nomoto, K., Yamaoka, H., Shigeyama, T., \& Iwamoto, K. 1996, in
        Supernovae and Supernova Remnants, ed. R. McCray \& Z. Wang
    (Cambridge: Cambridge Univ. Press), p. 49}
\ref{O'Connell, R.W. 1986, in Spectral Evolution of Galaxies,
     ed. C.Chiosi, A.Renzini, (Dordrecht: Reidel), p. 321}
\ref{Ohashi, T., 1995, Ann. N.Y. Aca. Sciences, 759, 217}
\ref{--------------.1996, MPE-Report, in press}
\ref{Ohasi, T., et al. 1996, PASJ, 48, 157} 
\ref{Ohashi, T., \& Tsuru, T. 1992, in Frontiers of X-Ray Astronomy, ed. Y.
     Tanaka \& K. Koyama (Tokyo: Universal Academy Press), p. 435}
\ref{Ohashi, T., et al. 1990, in Windows on Galaxies, ed. G. Fabbiano, J.
     Gallagher \& A. Renzini (Dordrecht: Kluwer), p. 243}
\ref{Peletier, R.F., Davies, R.L., Illingworth, G.D., Davis, L., Cawson,
     M., 1990, AJ, 100, 1091}
\ref{Pellegrini, S. 1994, A\&A, 292, 395}
\ref{Pellegrini, S., \& Fabbiano, G. 1994, ApJ, 429, 105}
\ref{Ponman, T.J., et al. 1994, Nature, 369, 462}
\ref{Raymond, J.C., \& Smith, B.W., 1977, ApJS, 35, 419}
\ref{Renzini, A. 1994a, in Panchromatic View of Galaxies, ed. G. Hensler et al.
    (Gif-sur-Yvette: Edition Fronti\'eres), p. 155}
\ref{--------------. 1994b, in Clusters of Galaxies, ed. F.
    Durret et al.(Gif-sur-Yvette: Edition Fronti\'eres), p. 221}
\ref{--------------. 1996a, in Supernovae and Supernova Remnants, ed.
    R. McCray \& Z. Wang (Cambridge: Cambridge Univ. Press), p. 77}
\ref{--------------. 1996b, in New Light on Galaxy Evolution, ed. R.
     Bender \& R. Davies (Dordrecht: Kluwer), in press}
\ref{--------------. 1996c, in preparation}
\ref{Renzini, A., Ciotti, L., D'Ercole, A., \& Pellegrini, S. 1993, ApJ,
    419, 52, RCDP}
\ref{Ruiz-Lapuente,P., Burkert, A., \& Canal, R. 1996,  ApJ 447, L69}
\ref{Sarazin, C.L. 1986, Rev. Mod. Phys. 58, 1}
\ref{Sakima, Y., Tawara, Y., \& Yamashita, K., 1994, in New Horizon of
     X-Ray Astronomy -- First Results From ASCA, 
     (Tokyo: Universal Academy Press), p. 557}
\ref{Serlemitos, P.J., Loewenstein, M., Mushotzky, R.F., Marshall, F.E,
    \& Petre, R. 1993, ApJ, 413, 518}
\ref{Shigeyama, Y., Nomoto, K., Yamaoka, H., \& Thielemann, F.-K. 1992, ApJ,
     386, L13}
\ref{Singh, K.P., White, N.E., \& Drake, S.A. 1996, ApJ, 456, 766}
\ref{Sneden, C., Lambert, D.L., \& Whitaker, R.W. 1979, ApJ, 234, 964}
\ref{Takano, S. 1989, PhD Thesis, University of Tokyo}
\ref{Tammann, G. 1982, in Supernovae: A Survey of Current Research, ed. M.
    Rees \& R. Stoneham (Dordrecht: Reidel), p. 371}
\ref{Tanaka, Y., Inoue, H., \& Holt, S.S., 1994,  PASJ, 46, L37} 
\ref{Thomsen, B., \& Baum, W.A. 1988, ApJ, 315, 460}
\ref{Tinsley, B.M. 1980, Fundam. Cosmic Phys. 5, 287}
\ref{Timmes, F.X., Woosley, S.E., \& Weaver, T.A. 1995, ApJS, 98, 617}
\ref{Trinchieri, G., Kim, D.-W., Fabbiano, G., \& Canizares, C.R., 1994, ApJ, 428, 555}
\ref{Tsujimoto, T., Nomoto, K., Hashimoto, M., Yanagida, S., Thielemann, F.-K.
     1993, in Nuclei in the Cosmos, ed. F. K\"appeler \& K. Wisshak (Bristol:
     IOP Publishing), p. 479}
\ref{Tsuru, T.  1993, PhD Thesis, University of Tokyo, ISAS RN 528}
\ref{Tsuru, T. et al. 1994, in New Horizon of X-ray Astronomy,
     ed. F.Makino, T.Ohashi, (Tokyo: Universal Academy Press), p. 529}
\ref{Tully, R.B., 1988, Nearby Calaxies Catalog, 
     Cambridge: Cambridge University Press)}
\ref{Ueno, S. et al. 1994, PASJ, 46, L71}
\ref{van den Bergh, S., \& Tammann, G. 1991, ARA\&A, 29, 363}
\ref{Wheeler, J.C., Sneden, C., \& Truran, J.W. 1989, ARA\&A, 27, 279}
\ref{White, N.E., 1996, in Cool Stars, Stellar Systems and the Sun,
     ed. N. Pallavicini, A.K. Dupree, PASP Conf. Ser., in press}
\ref{White, R.E.III, \& Sarazin, C.L. 1991, ApJ, 367, 476}
\ref{White, S.D.M., Navarro, J.F., Evrard, A.E., \& Frenk, C.S. 1993, Nature,
     366, 429}
\ref{Woosley, S.E., \& Weaver, T.A. 1994, in Les Houches, Session LIV,
     Supernovae, ed. S.R.Bludman, R.Mochkovitch, J.Zinn-Justin
     (Elsevier Science Publ.), p. 63}
\ref{Worthey, G. 1994, ApJS, 95, 107}
\ref{Worthey, G., Faber, S.M., \& Gonz\'alez, J.J. 1992, ApJ, 398, 69}

\vfill\eject

\centerline {\bf FIGURE CAPTION}

{\bf Figure 1:} Histrogram of the correction factor $\beta(c)$ 
calculated for our sample of elliptical galaxies listed in Table 1.
The luminosity-weighted iron abundance of a whole galaxy is given by
the product of  $\beta(c)$ times the luminosity-weighted abundance as
measured at the galaxy effective radius.
Two galaxies with $\log \beta(c)$ greater than 0.4 are NGC1453 and
NGC3962.

{\bf Figure 2:} Iron abundance of the ISM for the seven elliptical
galaxies in our sample (cf. Table 2) determined using the Raymond-Smith
model, as a function of \lx/\lb, where \lx is measured for $0.5-10$ keV
and given in units of erg s$^{-1}$ and \lb is in solar units. 
Open asterisks indicate the mean stellar iron abundances (cf. Table 1).

{\bf Figure 3:} Iron abundance of the ISM for the same galaxies as in
Fig. 2 as a function of temperature of the ISM, also
determined using the Raymond-Smith model.

{\bf Figure 4:} The iron abundances of the ISM of the same seven
galaxies in Fig. 2 and 3, as a function of the ISM temperature. 
Solid, solid-thick, and dotted error bars  indicate
values obtained by the Raymond-Smith model, Meka model, and
Masai model, respectively. Indivisual galaxies are identified by the
same symbol at the center of the error bars.

{\bf Figure 5:} The iron abundance of  the $\,$ ICM of clusters
and groups as a function of the ICM temperature.
Data are taken from the following sources: Filled circles: Arnaud
et al. (1992); filled triangles: Tsuru (1993); open triangle: David   et al.
(1994a); open square: Mulchaey et al. (1993); filled square:
Ponman et al. (1994); open circles: Mulchaey et al. (1996). For
$kT\gsim 3$ keV the iron abundance is derived from the iron-K complex,
while for lower temperatures the iron-L complex is used.

\vfill\eject

\baselineskip=5truemm
\centerline {}
\centerline {\bf TABLE 1}
\centerline {\bf MEAN STELLAR IRON ABUNDANCES OF ELLIPTICAL GALAXIES}
\vskip 0.5cm
\hrule
\vskip 1truemm
\hrule
\vskip 3 truemm
\settabs\+ NGC12345 & 99 \quad & ref \quad &  Mg$_2$(0)
&  1234567 \quad & -9.999 & 99 \quad & [Fe/H]$_0$ 
& 12345678 \quad & 999 \quad & 99 \quad & 999 \quad \cr

\+ Galaxy & $r_e$ & ref. & Mg$_2$(0) & Mg$_2$($r_e$) & $d$ & ref. & 
[Fe/H]$_0$ & $<$[Fe/H]$>$ & $\sigma_0$ \cr
\+ \hskip 0.4cm (1) & (2) & (3) & \hskip 0.2cm (4) & \hskip 0.2cm (5)
& \hskip -0.1cm (6) 
& (7) & \hskip 0.3cm (8) & \hskip 0.5cm (9) & (10) \cr  
\vskip 0.3cm
\hrule
\vskip 0.3cm
 
\+ NGC315  & 32 & \ 2  & 0.283 & 0.282 & \hskip -0.3cm --0.031 & \ 2 &\hskip 0.2cm 0.03 & \hskip 0.5cm 0.05   &  352\cr
\+ NGC439  & 45 & 11   &       & 0.231 & \hskip -0.3cm --0.053 & 11 &\hskip 0.2cm      & \hskip 0.4cm --0.27   &     \cr
\+ NGC741  & 50 & \ 1  & 0.324 & 0.234 & \hskip -0.3cm --0.057 & \ 5 &\hskip 0.2cm 0.34 & \hskip 0.4cm --0.23 &  280 \cr
\+ NGC741  & 42 & \ 2  & 0.324 & 0.240 & \hskip -0.3cm --0.058 & \ 2 & \hskip 0.2cm 0.34 & \hskip 0.4cm --0.18 & 280 \cr
\+ NGC1052 & 30 & \ 4  & 0.316 & 0.261 & \hskip -0.3cm --0.061 & \ 4 & \hskip 0.2cm 0.28 & \hskip 0.4cm --0.02 & 206 \cr
\smallskip

\+ NGC1453 & 30 & \ 1  & 0.327 & 0.168 & \hskip -0.3cm --0.118 & \ 5 & \hskip 0.2cm 0.36 & \hskip 0.4cm --0.37 & 290 \cr
\+ NGC1600 & 34 & \ 2  & 0.324 & 0.257 & \hskip -0.3cm --0.088 & \ 2 & \hskip 0.2cm 0.34 & \hskip 0.5cm  0.08 & 321 \cr
\+ NGC2434 & 24 & 11   & 0.268 & 0.187 & \hskip -0.3cm --0.040 & 11  & \hskip 0.2cm      & \hskip 0.4cm --0.63 & 205 \cr
\+ NGC2663 & 50 & 10   & 0.324 & 0.269 & \hskip -0.3cm --0.051 & 10  & \hskip 0.2cm      & \hskip 0.5cm  0.01 & 281 \cr
\+ NGC2693 & 17 & \ 1  & 0.328 & 0.273 & \hskip -0.3cm --0.041 & \ 5 & \hskip 0.2cm 0.37 & \hskip 0.5cm  0.01 & 279 \cr
\smallskip

\+ NGC3379 & 45 & \ 1  & 0.308 & 0.204 & \hskip -0.3cm --0.073 & \ 5 & \hskip 0.2cm 0.22 & \hskip 0.4cm --0.39 & 201 \cr
\+ NGC3379 & 56 & \ 2  & 0.308 & 0.238 & \hskip -0.3cm --0.065 & \ 2 & \hskip 0.2cm 0.22 & \hskip 0.4cm --0.17 & 201 \cr
\+ NGC3379 & 56 & \ 4  & 0.308 & 0.261 & \hskip -0.3cm --0.064 & \ 4 & \hskip 0.2cm 0.22 & \hskip 0.5cm  0.00 & 201 \cr
\+ NGC3665 & 34 & \ 2  & 0.296 & 0.254 & \hskip -0.3cm --0.022 & \ 2 & \hskip 0.2cm 0.12 & \hskip 0.4cm --0.17 & 205 \cr
\+ NGC3706 & 27 & 11   & 0.310 & 0.216 & \hskip -0.3cm --0.062 & 11  & \hskip 0.2cm      & \hskip 0.4cm --0.35 &     \cr
\smallskip

\+ NGC3818 & 19 & \ 3  & 0.315 & 0.236 & \hskip -0.3cm --0.083 & \ 3 & \hskip 0.2cm 0.27 & \hskip 0.4cm --0.10 & 206 \cr
\+ NGC3904 & 19 & \ 3  & 0.312 & 0.228 & \hskip -0.3cm --0.084 & \ 3 & \hskip 0.2cm 0.24 & \hskip 0.4cm --0.15 & 215 \cr
\+ NGC3962 & 33 & \ 1  & 0.306 & 0.129 & \hskip -0.3cm --0.146 & \ 5 & \hskip 0.2cm 0.20 & \hskip 0.4cm --0.38 & 211 \cr
\+ NGC4261 & 41 & \ 2  & 0.330 & 0.271 & \hskip -0.3cm --0.068 & \ 2 & \hskip 0.2cm 0.38 & \hskip 0.5cm  0.09 & 294 \cr
\+ NGC4278 & 39 & \ 1  & 0.291 & 0.223 & \hskip -0.3cm --0.066 & \ 5 & \hskip 0.2cm 0.09 & \hskip 0.4cm --0.28 & 266 \cr
\smallskip
 
\+ NGC4278 & 30 & \ 2  & 0.291 & 0.234 & \hskip -0.3cm --0.088 & \ 2 & \hskip 0.2cm 0.09 & \hskip 0.4cm --0.09 & 266 \cr
\+ NGC4283 &  9 & \ 1  & 0.268 & 0.233 & \hskip -0.3cm --0.052 & \ 5 &\hskip 0.1cm --0.08 & \hskip 0.4cm --0.26 & 100 \cr
\+ NGC4365 & 53 & \ 4  & 0.300 & 0.263 & \hskip -0.3cm --0.066 & \ 4 & \hskip 0.2cm 0.15 & \hskip 0.5cm  0.02 & 248 \cr
\+ NGC4374 & 53 & \ 2  & 0.305 & 0.262 & \hskip -0.3cm --0.050 & \ 2 & \hskip 0.2cm 0.19 & \hskip 0.4cm --0.05 & 287 \cr
\+ NGC4406 & 70 & \ 4  & 0.311 & 0.257 & \hskip -0.3cm --0.031 & \ 4 & \hskip 0.2cm 0.23 & \hskip 0.4cm --0.13 & 250 \cr
\smallskip

\+ NGC4472 & 99 & \ 2  & 0.306 & 0.297 & \hskip -0.3cm --0.028 & \ 2 & \hskip 0.2cm 0.20 & \hskip 0.5cm  0.16 & 287 \cr
\+ NGC4472 & 99 & \ 3  & 0.306 & 0.275 & \hskip -0.3cm --0.042 & \ 3 & \hskip 0.2cm 0.20 & \hskip 0.5cm  0.03 & 287 \cr
\+ NGC4478 & 15 & \ 3  & 0.253 & 0.229 & \hskip -0.3cm --0.014 & \ 3 &\hskip 0.1cm --0.20 & \hskip 0.4cm --0.36 & 149 \cr
\+ NGC4486 & 76 & \ 1  & 0.289 & 0.177 & \hskip -0.3cm --0.125 & \ 5 & \hskip 0.2cm 0.07 & \hskip 0.4cm --0.24 & 361 \cr
\+ NGC4486 & 95 & \ 2  & 0.289 & 0.270 & \hskip -0.3cm --0.076 & \ 2 & \hskip 0.2cm 0.07 & \hskip 0.5cm  0.12 & 361 \cr
\smallskip

\+ NGC4621 & 42 & \ 4  & 0.328 & 0.265 & \hskip -0.3cm --0.034 & \ 4 & \hskip 0.2cm 0.37 & \hskip 0.4cm --0.07 & 240 \cr
\+ NGC4636 & 67 & \ 2  & 0.311 & 0.236 & \hskip -0.3cm --0.077 & \ 2 & \hskip 0.2cm 0.23 & \hskip 0.4cm --0.13 & 191 \cr
\+ NGC4742 & 23 & \ 3  & 0.177 & 0.175 & \hskip -0.3cm --0.030 & \ 3 &\hskip 0.1cm --0.76 & \hskip 0.4cm --0.74 &  93 \cr
\+ NGC4839 & 42 & \ 2  & 0.315 & 0.246 & \hskip -0.3cm --0.058 & \ 2 & \hskip 0.2cm 0.26 & \hskip 0.4cm --0.14 & 259 \cr
\+ NGC4874 & 67 & \ 1  & 0.328 & 0.219 & \hskip -0.3cm -0.060 & \ 6 & \hskip 0.2cm 0.37 & \hskip 0.4cm --0.33 & 245 \cr
\vskip 0.3cm
\hrule
\vfill\eject

\centerline {\bf TABLE 1  (continued)}
\vskip 0.5cm
\hrule
\vskip 1truemm
\hrule
\vskip 3 truemm
\+ Galaxy & $r_e$ & ref. & Mg$_2$(0) & Mg$_2$($r_e$) & $d$ & ref. & 
[Fe/H]$_0$ & $<$[Fe/H]$>$ & $\sigma_0$ \cr
\+ \hskip 0.4cm (1) & (2) & (3) & \hskip 0.2cm (4) & \hskip 0.2cm (5) & \hskip -0.1cm (6) 
& (7) & \hskip 0.3cm (8) & \hskip 0.5cm (9) & (10) \cr  

\vskip 0.3cm
\hrule
\vskip 0.3cm
\+ NGC4881 & 11 & \ 1  & 0.293 & 0.225 & \hskip -0.3cm --0.088 & \ 7 & \hskip 0.2cm 0.10 & \hskip 0.4cm --0.15 & 219 \cr
\+ NGC4915 &  9 & \ 1  & 0.291 & 0.230 & \hskip -0.3cm --0.044 & \ 5 & \hskip 0.2cm 0.09 & \hskip 0.4cm --0.30 & 209 \cr
\+ NGC5018 & 22 & 10   & 0.209 & 0.209 & \hskip -0.3cm --0.034 & 10  & \hskip 0.2cm      & \hskip 0.4cm --0.48 & 223 \cr
\+ NGC5638 & 24 & \ 3  & 0.317 & 0.262 & \hskip -0.3cm --0.040 & \ 3 & \hskip 0.2cm 0.28 & \hskip 0.4cm --0.07 & 159 \cr
\+ NGC5813 & 44 & \ 3  & 0.308 & 0.157 & \hskip -0.3cm --0.120 & \ 8 & \hskip 0.2cm 0.21 & \hskip 0.4cm --0.43 & 238 \cr
\smallskip

\+ NGC5813 & 44 & \ 3  & 0.308 & 0.175 & \hskip -0.3cm --0.089 & \ 3 & \hskip 0.2cm 0.21 & \hskip 0.4cm --0.52 & 238 \cr
\+ NGC5831 & 27 & \ 3  & 0.289 & 0.206 & \hskip -0.3cm --0.060 & \ 3 & \hskip 0.2cm 0.07 & \hskip 0.4cm --0.43 & 166 \cr
\+ NGC5845 &  8 & \ 3  & 0.304 & 0.270 & \hskip -0.3cm --0.049 & \ 3 & \hskip 0.2cm 0.18 & \hskip 0.5cm  0.01 & 251 \cr
\+ NGC6407 & 33 & 11   &       & 0.255 & \hskip -0.3cm --0.035 & 11  & \hskip 0.2cm      & \hskip 0.4cm --0.14 &     \cr
\+ NGC7192 & 28 & 11   & 0.250 & 0.188 & \hskip -0.3cm --0.084 & 11  & \hskip 0.2cm      & \hskip 0.4cm --0.45 & 185 \cr
\smallskip

\+ NGC7332 & 17 & \ 4  & 0.278 & 0.243 & \hskip -0.3cm --0.061 & \ 4 &\hskip 0.1cm --0.01 & \hskip 0.4cm --0.15 &     \cr
\+ NGC7626 & 34 & \ 2  & 0.336 & 0.262 & \hskip -0.3cm --0.063 & \ 2 & \hskip 0.2cm 0.42 & \hskip 0.5cm  0.00 & 234 \cr
\+ IC 4296 & 48 & \ 3  & 0.323 & 0.200 & \hskip -0.3cm --0.087 & \ 3 & \hskip 0.2cm 0.32 & \hskip 0.4cm --0.35 & 323 \cr
\+ 2354-35 & 80 & \ 3  & 0.285 & 0.248 & \hskip -0.3cm --0.021 & \ 3 & \hskip 0.2cm 0.04 & \hskip 0.4cm --0.22 & 316 \cr
\vskip 0.3cm
\hrule
\vskip 1.5cm

  \pn
  References to columns (3) and (7): 

\noindent  1) Kodaira et al. (1990), 2) Davies et al. (1993), 
  3) Gorgas et al. (1990), 4) Couture \& Hardy (1988), 5) Davidge (1992), 
  6) Baum et al. (1986), 7) Thomsen \& Baum  (1987), 
  8) Efstathiou \& Gorgas (1985), 9) Davies et al. (1987), 
  10) Carollo \& Danziger (1994a), 11) Carollo \& Danziger (1994b)

\vfill\eject

\baselineskip=7truemm
\centerline {\bf TABLE 2}
\centerline {\bf IRON  IN THE HOT ISM OF ELLIPTICALS FROM THE RS MODEL}
\vskip 0.5cm
\hrule
\vskip 1truemm
\hrule
\vskip 3 truemm
\settabs\+ NGC12345& \lb a9000aa & aaaaaaa& 0.69(0.50\quad &Abundance & nHaaaaaaaa  &Lx(erg/s)  \cr

\+ Galaxy     &log \lb & $\sigma_m$& kT             & Abundance      & nH &log Lx  \cr
\+            &(\lsun )      &(km/s)    & (keV)          & (solar)    & $(10^{21}\rm{cm^{-2})}$&(erg/s)\cr
\+  (1)       &(2)      &(3)       &(4)             &(5)             & (6)  &(7)   \cr
\vskip 0.2cm
\hrule
\vskip 0.2cm                     
\+ NGC720     &10.3  &247 &$0.65^{+0.04}_{-0.04}$&$0.14^{+0.07}_{-0.05}$&$0.3^{+0.7}_{-0.3}$&40.6 & \cr
\+ NGC1399    &10.3  &310 &$1.05^{+0.01}_{-0.02}$&$0.42^{+0.08}_{-0.07}$&$1.1^{+0.0}_{-0.2}$&41.7$^*$ \cr 
\+ NGC1404    &10.2  &225 &$0.66^{+0.03}_{-0.03}$&$0.18^{+0.09}_{-0.05}$&$0.6^{+0.4}_{-0.4}$&40.7\cr 
\+ NGC4374    &10.6  &287 &$0.68^{+0.06}_{-0.06}$&$0.11^{+0.07}_{-0.05}$&$1.4^{+0.6}_{-0.7}$&40.8   \cr 
\+ NGC4406    &10.7  &250 &$0.83^{+0.01}_{-0.01}$&$0.27^{+0.10}_{-0.04}$&$0.9^{+0.2}_{-0.2}$&42.0\cr 
\+ NGC4472    &10.9  &287 &$0.93^{+0.02}_{-0.02}$&$0.33^{+0.07}_{-0.05}$&$1.1^{+0.2}_{-0.2}$&41.8   \cr 
\+ NGC4636    &10.5  &191 &$0.74^{+0.02}_{-0.01}$&$0.28^{+0.07}_{-0.04}$&$0.9^{+0.2}_{-0.1}$&41.7\cr 
\vskip 3 truemm
\hrule
\vskip 1cm

  Notes to Table 2:

  Column (1): galaxy ID

  Column (2): B-luminosity (Tully 1988)

  Column (3): mean velocity dispersion (Davies et al. 1987)

  Column (4): temperature

  Column (5): iron abundance (solar value = $4.68\ 10^{-5}$)

  Column (6): hydrogen column density 

  Column (7): X-ray luminosity (0.5-10 keV)

  *) The X-ray luminosity for NGC1399 is for r $< $ 10 arcmin

\vfill\eject

\baselineskip=7truemm
\centerline {\bf TABLE 3}
\centerline {\bf IRON  IN THE HOT ISM OF ELLIPTICALS
FROM THE MEKA AND MASAI MODELS}
\vskip 0.5cm
\hrule
\vskip 1truemm
\hrule
\vskip 3 truemm
\settabs\+ NGC12345 & 0.69(0.50-0.\quad & Abundance &nHaaaaaa &0.69(0.50-0.\quad & Abundance & nHaaaaaaa \cr

\+ Model  & & Meka &       & & Masai &      &  \cr
\+ Galaxy & kT & Abundance & nH &kT & Abundance &nH     \cr
\+        & (keV)& (solar) &  $(10^{21}\rm{cm^{-2}})$   &(keV)& (solar) &  $(10^{21}\rm{cm^{-2}})$   \cr
\+  (1)       &(2)      &(3)       &(4)             &(5)             & (6)  &(7)   \cr
\vskip 0.3cm
\hrule
\vskip 0.3cm
\+ NGC720   &$0.43^{+0.06}_{-0.05}$&$0.07^{+0.22}_{-0.04}$&$1.3^{+1.1}_{-1.3}$  &
             $0.42^{+0.20}_{-0.14}$&$0.10^{+0.30}_{-0.05}$&$1.0^{+1.0}_{-1.0}$ \cr
\+ NGC1399  &$0.82^{+0.02}_{-0.02}$&$0.27^{+0.05}_{-0.04}$&$1.5^{+0.2}_{-0.2}$&
             $0.98^{+0.02}_{-0.02}$&$0.58^{+0.16}_{-0.08}$&$1.1^{+0.2}_{-0.2}$ \cr
\+ NGC1404  &$0.46^{+0.05}_{-0.06}$&$0.16^{+0.80}_{-0.08}$&$1.4^{+1.0}_{-1.4}$&
             $0.43^{+0.10}_{-0.05}$&$0.20^{+0.20}_{-0.10}$&$1.6^{+1.5}_{-0.6}$\cr
\+ NGC4374  &$0.56^{+0.06}_{-0.05}$&$0.15^{+0.25}_{-0.08}$&$1.5^{+0.8}_{-0.7}$&
             $0.67^{+0.06}_{-0.06}$&$0.22^{+0.30}_{-0.13}$&$1.1^{+0.6}_{-0.6}$\cr
\+ NGC4406  &$0.66^{+0.01}_{-0.01}$&$0.37^{+0.07}_{-0.05}$&$1.0^{+0.1}_{-0.2}$ &
             $0.79^{+0.02}_{-0.01}$&$0.64^{+0.20}_{-0.10}$&$0.6^{+0.2}_{-0.2}$\cr
\+ NGC4472  &$0.73^{+0.01}_{-0.01}$&$0.32^{+0.06}_{-0.07}$&$1.5^{+0.2}_{-0.2}$ &
             $0.86^{+0.01}_{-0.01}$&$0.75^{+0.20}_{-0.15}$&$0.7^{+0.2}_{-0.2}$\cr
\+ NGC4636  &$0.58^{+0.01}_{-0.03}$&$0.40^{+0.15}_{-0.10}$&$1.1^{+0.2}_{-0.2}$&
             $0.70^{+0.01}_{-0.02}$&$0.82^{+0.46}_{-0.23}$&$0.5^{+0.2}_{-0.3}$\cr

\vskip 0.3cm
\hrule
\vskip 1cm

  Notes to Table 3:

  Column (1): galaxy ID

  Column (2): temperature (Meka model)

  Column (3): iron abundance (Meka model)

  Column (4): hydrogen column density (Meka model)

  Column (5): temperature (Masai model)

  Column (6): iron abundance (Masai model)

  Column (7): hydrogen column density (Masai model)

\vfill \eject
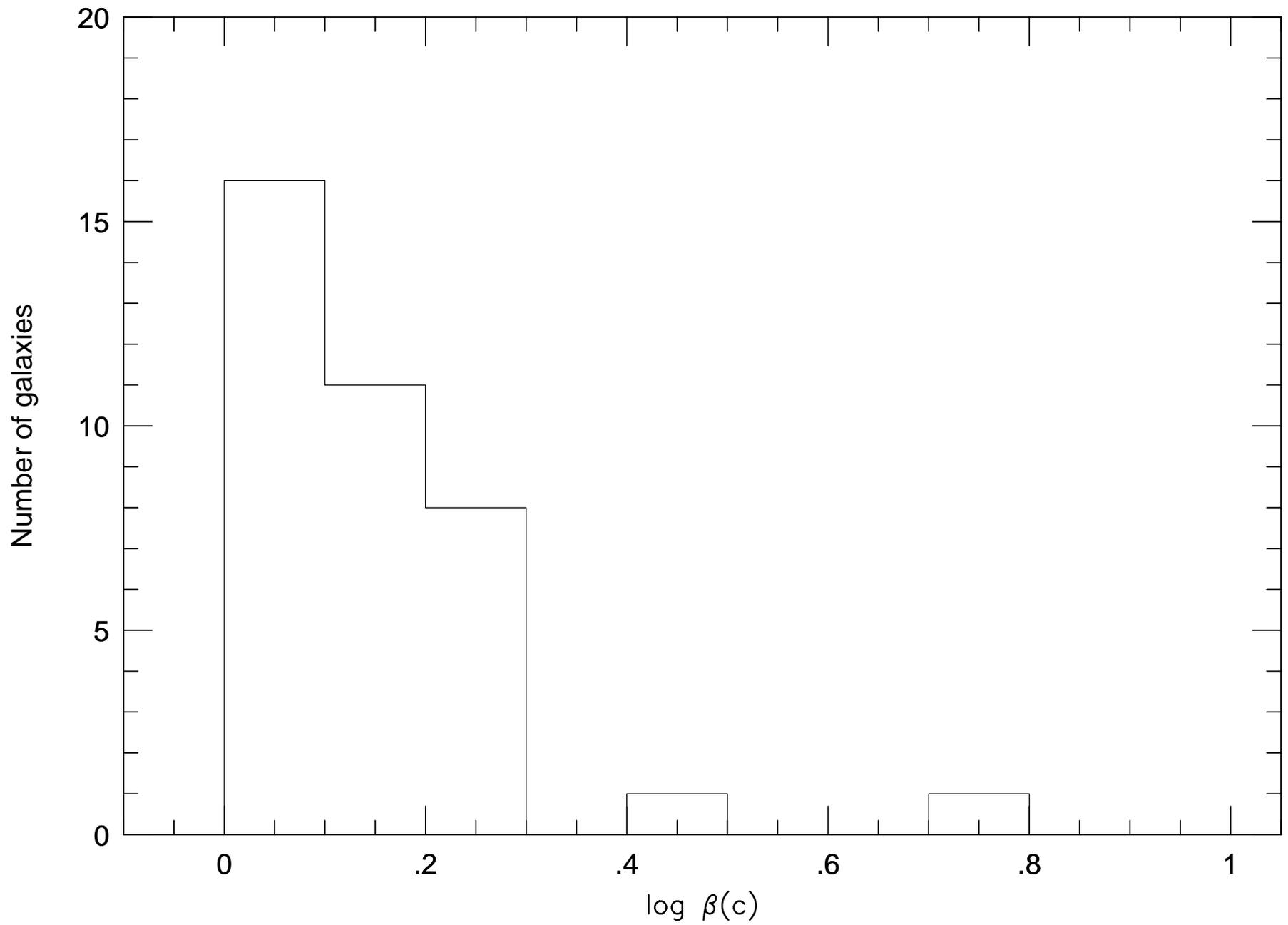
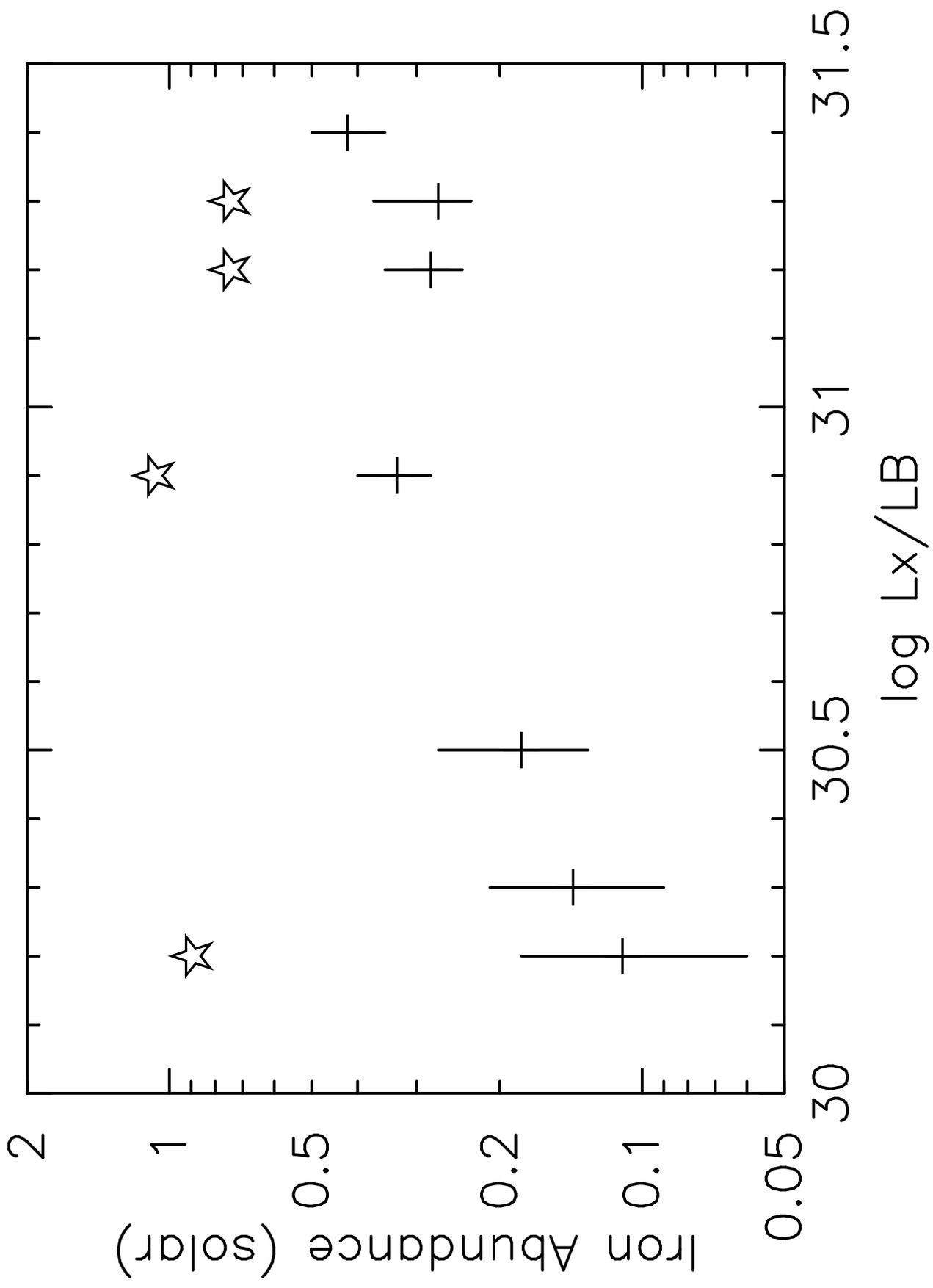
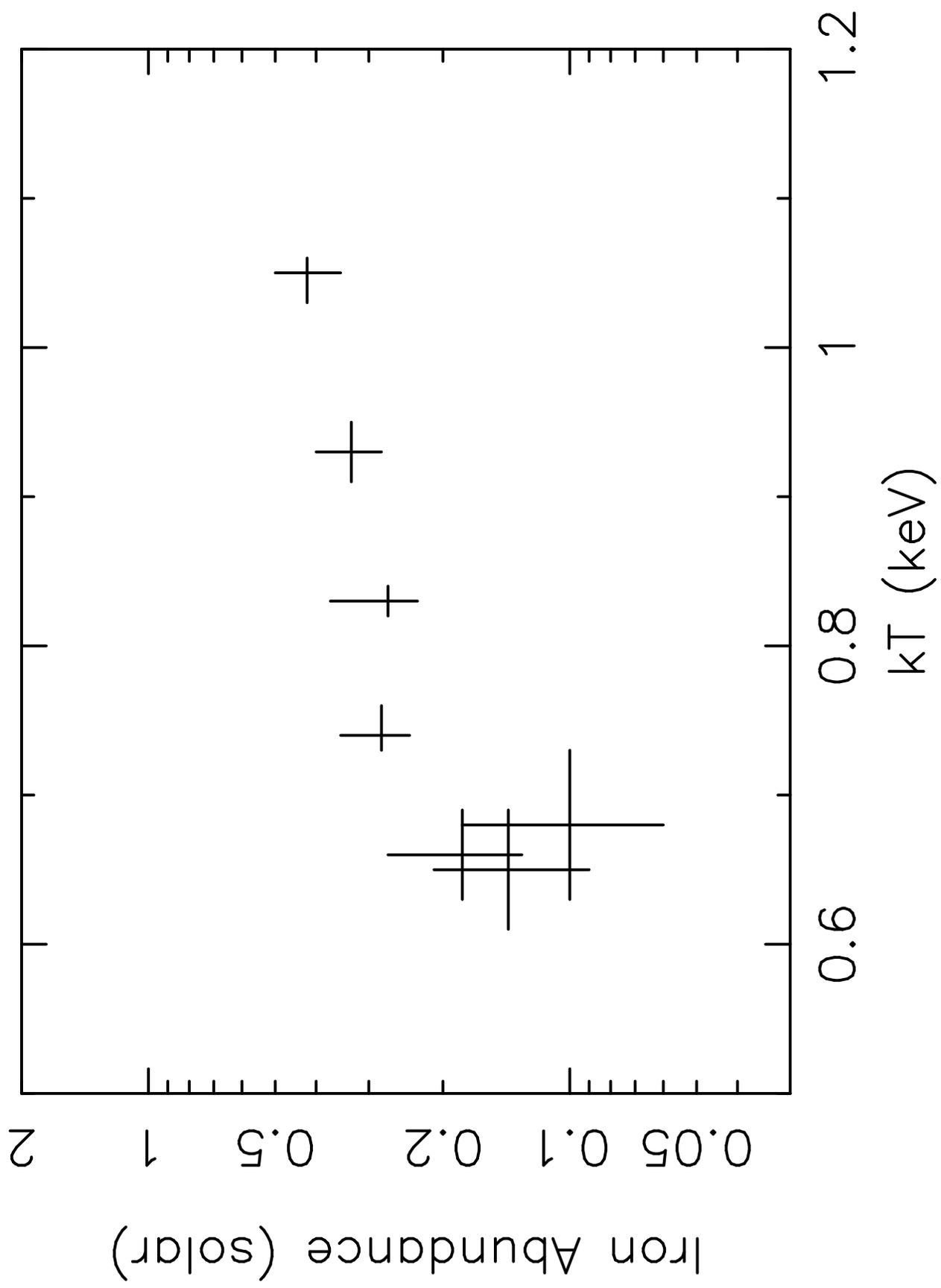
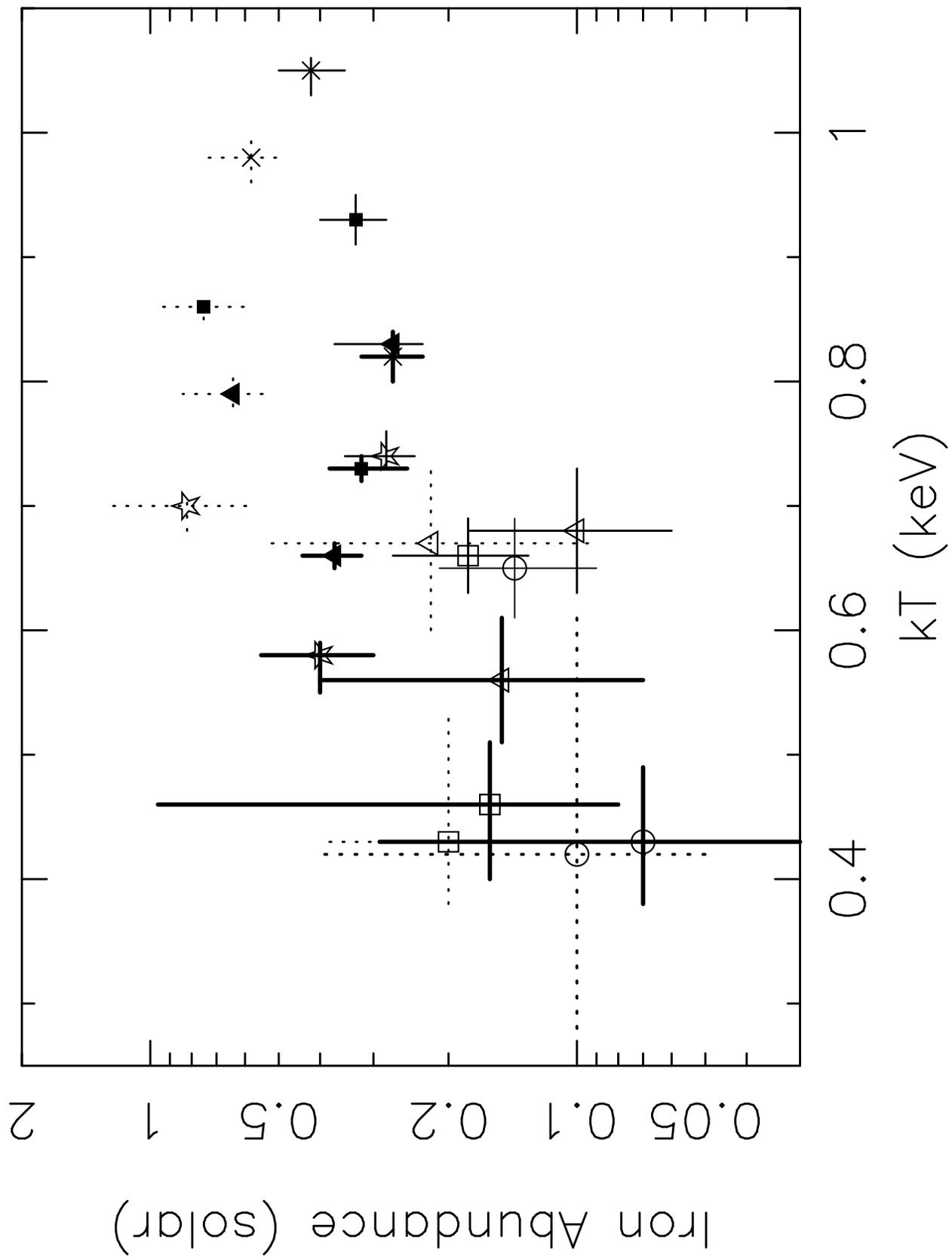
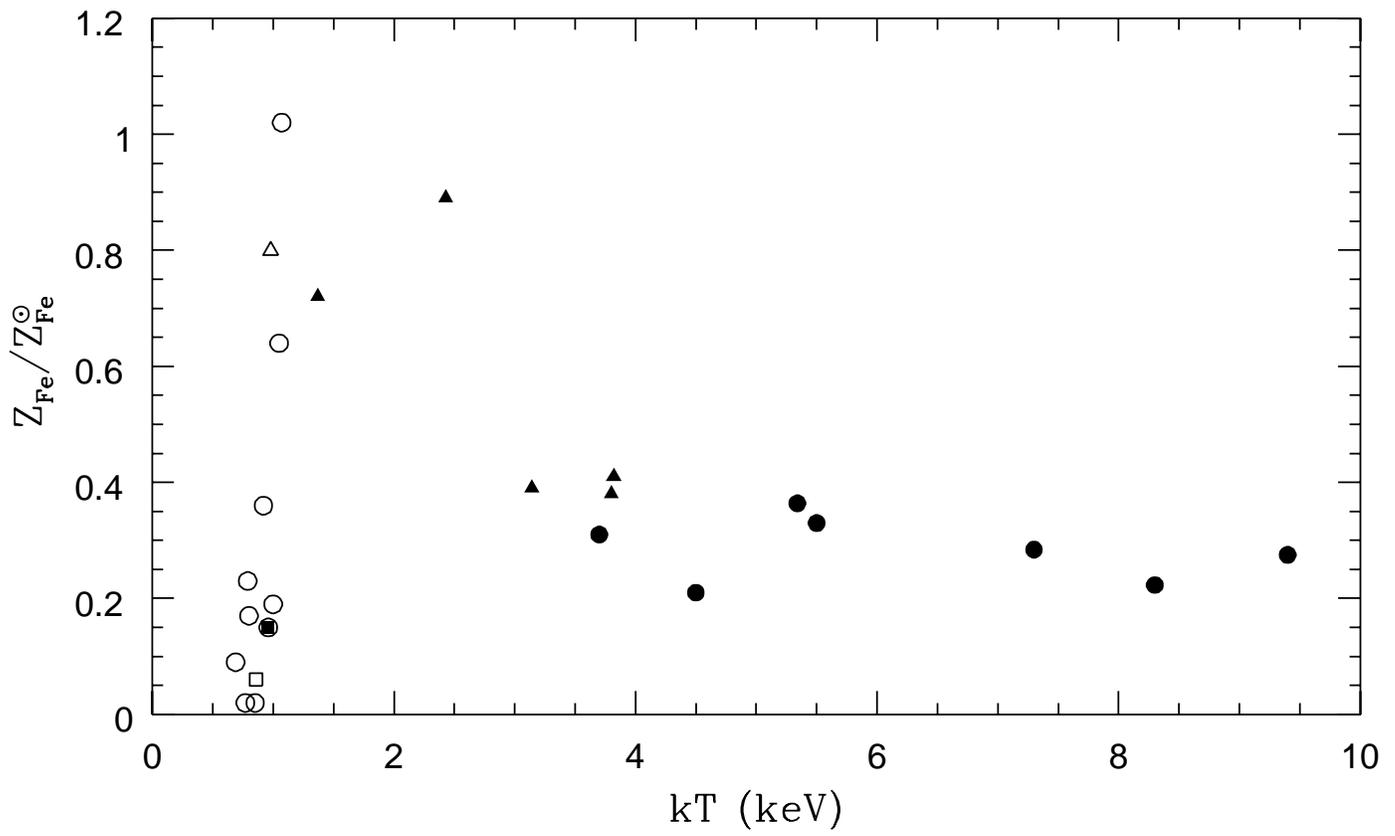
\end